\documentclass[useAMS]{mn2e}

\usepackage{times}
\usepackage{graphicx}
\usepackage{amsmath}
\usepackage{amssymb}
\newcommand{\beq}{\begin{equation}}
\newcommand{\bea}{\begin{eqnarray}}
\newcommand{\eeq}{\end{equation}}
\newcommand{\eea}{\end{eqnarray}}

\newcommand{\bfs}[1]{\mbox{\boldmath{$#1$}}}

\title[Instabilities at relativistic shock waves]{Dispersion and
  thermal effects on electromagnetic instabilities in the precursor of
  relativistic shocks}

\author[M. Lemoine and G. Pelletier]
{Martin Lemoine$^{1}$\thanks{e-mail:{\tt lemoine@iap.fr}} and
Guy Pelletier$^{2}$\thanks{e-mail:{\tt
     guy.pelletier@obs.ujf-grenoble.fr}}\\
$^{1}$ Institut d'Astrophysique de Paris, \\ 
	CNRS, UPMC, 
	98 bis boulevard Arago, F-75014 Paris, France\\
$^{2}$ Laboratoire d'Astrophysique de Grenoble, \\
	CNRS, Universit\'e Joseph Fourier II,
	BP 53, F-38041 Grenoble, France; \\
}

\begin{document}

\date{}

\pubyear{2008}

\maketitle

\label{firstpage}

\begin{abstract}
  Fermi acceleration can develop efficiently at relativistic
  collisionless shock waves provided the upstream (unshocked) plasma
  is weakly magnetized.  This has been both indicated by analytical
  theory and observed in numerical particle-in-cell simulations. At
  low magnetization, the large size of the shock precursor indeed
  provides enough time for electromagnetic micro-instabilities to grow
  and such micro-instabilities generate small scale turbulence that in
  turn provides the scattering required for particles to undergo Fermi
  cycles at superluminal relativistic shock waves. The present paper
  extends our previous analysis on the development of these
  micro-instabilities to account for the finite angular dispersion of
  the beam of reflected and accelerated particles and to account for
  the expected heating of the upstream electrons in the shock
  precursor. Indeed we argue that the electrons can be significantly
  heated during their travel in the shock precursor and that they may
  even reach equipartition with protons, in agreement with recent
  numerical simulations. We show that the oblique two stream
  instability may operate down to values of the shock Lorentz factor
  $\gamma_{\rm sh}\sim 10$ (corresponding to a relatively large
  angular dispersion of the beam) as long as the electrons of the
  upstream plasma remain cold, while the filamentation instability is
  strongly inhibited in this limit; however, as electrons get heated
  to relativistic temperatures, the situation becomes opposite and the
  two stream instability becomes inhibited while the filamentation
  mode becomes efficient, even at moderate values of the shock Lorentz
  factor. The peak wavelength of these instabilities migrates from the
  inertial electron scale towards the proton inertial scale as the
  background electrons get progressively heated during the crossing of
  the shock precursor. We also discuss the emergence and the role of
  current driven instabilities upstream of the shock. In particular,
  we show that the returning and accelerated particles give rise to a
  transverse current through their rotation in the background magnetic
  field. We find that the compensating current in the background
  plasma can lead to a Buneman instability which provides an efficient
  source of electron heating.
\end{abstract}


\section{Introduction}
The acceleration of particles through repeated interactions with the
electromagnetic fields up- and down-stream of a collisionless shock
front is generally taken as the source the non-thermal particle
populations that are detected in powerful astrophysical outflows. The
physics of this Fermi mechanism is relatively well understood for
non-relativistic shock velocities, at least in the test particle limit
in which one neglects the backreaction of the accelerated particles on
the shock environment. In the relativistic limit however, in which the
shock moves toward the upstream (unshocked) medium with a bulk Lorentz
factor $\gamma_{\rm sh}\,\gg\,1$, the situation becomes more
intricate, mostly because the shock now moves about as fast as the
accelerated particle. In non-relativistic shocks one can distinguish
the details of the shock structure from that of the Fermi process, as
this latter takes place on spatial scales much larger than the shock
thickness. On the contrary, in relativistic shocks, the structure of
the shock plays a significant role in the Fermi process and in the
process of generation of the electromagnetic turbulence. One important
issue in particular, is the reflection of a part of the cold upstream
particles at the shock front, which together with the Fermi
accelerated particles, travel back upstream and initiate instabilities
in the upstream plasma.

These instabilities play a key role in the development of the Fermi
process and consequently, in the radiative signatures of relativistic
outflows. Indeed, at magnetized oblique ultra-relativistic shock
waves, the Fermi process may develop only if sufficient
micro-turbulence has been generated in order to prevent the advection
of the accelerated particles away from the shock front into the far
downstream plasma (Begelman \& Kirk 1990, Niemiec \& Ostrowski 2006,
Lemoine et al. 2006). And, for the vast majority of them,
ultra-relativistic shock waves are oblique: the opposite parallel
configuration only takes effect when the magnetic field lines are
oriented to an angle with the shock normal smaller than $1/\gamma_{\rm
  sh}$ (in the unshocked plasma or upstream rest frame). As to the
meaning of ``magnetized'' shock waves, it is slightly ambiguous and
context dependent; above, it is meant: with a pre-existing (upstream)
magnetic field $B_{\rm u}$ such that, after compression through the
shock and Lorentz transform into the downstream frame, the typical
Larmor radius of accelerated particles is much smaller than the width
of the blast, which is of the order of $R/\gamma_{\rm sh}$ in this
frame ($R$ denotes the shock radius). For smaller values of $B_{\rm
  u}$, the accelerated particles can explore the whole blast and
beyond, hence they may scatter on magnetic inhomogeneities sourced
elsewhere, which would allow their repeated interactions with the
shock. Nevertheless, for such low values of $B_{\rm u}$, the
generation of micro-turbulence is guaranteed, as will be discussed in
the following, hence the Fermi process should develop independently of
any extra source of magnetic inhomogeneities\footnote{More generally
  speaking, Fermi acceleration requires micro-turbulence to be
  generated at distances smaller than $r_{\rm L,0}$, with $r_{\rm
    L,0}$ the Larmor radius in the downstream frame shock compressed
  background magnetic field (Pelletier et al. 2009). While this is
  guaranteed for micro-instabilities when they can be triggered in the
  precursor, this constraint can be turned into a lower bound on the
  growth rate of extra sources of turbulence, which of course differs
  from the constraint on the growth of micro-instabilities discussed
  here and in Lemoine \& Pelletier (2010).}  Furthermore, there exist
evidence for amplification of the magnetic field upstream of a
relativistic shock, see e.g. Li \& Waxman (2006), Li \& Zhao (2011),
which can be seen as a hallmark of the development of
micro-instabilities upstream of the shock.

On microscopic plasma scales, several instabilities may develop in the
precursor of the shock front through the penetration of the beam of
returning and accelerated particles into the upstream plasma. The
filamentation instability in particular -- often referred to as the
Weibel instability in the literature -- has been proposed as the agent
of amplification of the magnetic field to the level inferred from the
synchrotron interpretation of gamma-ray burst afterglows (Gruzinov \&
Waxman 1999, Medvedev \& Loeb 1999; see also Wiersma \& Achterberg
2004, Lyubarsky \& Eichler 2006, Achterberg \& Wiersma 2007,
Achterberg et al. 2007). Recently, it has been found that the growth
rate of two-stream instabilities in an oblique configuration (Bret,
Firpo \& Deutsch 2005) is larger than that of the filamentation mode
(Bret 2009, Lemoine \& Pelletier 2010). Which instability grows faster
is a key question here, as the relativistic velocity of the shock wave
and the level of upstream magnetization strongly limit the penetration
length scale of the accelerated particles into the upstream plasma
before these particles are caught back by the shock wave
(Milosavljevi\'c \& Nakar 2006, Pelletier et al.  2009); and as
mentioned before, if such instabilities cannot grow, then Fermi
acceleration cannot develop (at least, in the absence of extra sources
of turbulence).

The detailed conditions under which the above instabilities and other
relevant modes can grow, together with the development of Fermi cycles
as a function of the shock velocity and the upstream magnetization
have been discussed in a previous paper (Lemoine \& Pelletier
2010). That paper assumed a charge neutralized beam without angular
dispersion propagating in a cold background plasma. It is the aim of
the present paper to revisit these assumptions and to extend the
previous calculations to a more generic situation, in which the beam
angular dispersion (which is fixed by kinematics, as discussed below)
is taken into account, and in which the effects of heating of the
upstream plasma are considered. The emergence of current instabilities
is also discussed. The present study is warranted in particular by
recent work (Lyubarsky \& Eichler 2006, Rabinak et al. 2011) that
showed that the small opening angle of the beam suffices to prevent
the onset of the filamentation instability in electron-proton plasmas
of shock Lorentz factor $\gamma_{\rm sh}\,\lesssim\,100$,
independently of the length scale of the precursor. The present study
recovers this result for a cold background plasma, but it also argues
that: (i) the two-stream instability and more precisely its
  oblique mode version is less sensitive to the beam angular
dispersion as it can develop down to $\gamma_{\rm sh}\sim 10$ in the
same conditions; (ii) the heating of the background plasma has a major
effect in that it may help sustain the development of instabilities
down to $\gamma_{\rm sh}\sim 10$.

The present paper is laid out as follows. In Section~\ref{sec:prelim},
we first review the current understanding on the structure of the
shock, relying on the results of the most recent particle-in-cell
(PIC) simulations of relativistic collisionless shocks. We also
discuss in a general way the instabilities that may develop as a
function of the two main characteristics of the shock: upstream
magnetization and shock velocity. In Section~\ref{sec:coldback}, we
then discuss how the filamentation and two-stream instabilities are
affected by the finite angular dispersion of the beam of returning
particles, assuming the background plasma to remain cold. In
Section~\ref{sec:hotback} we generalize the calculations of
Sec.~\ref{sec:coldback} to a background plasma composed of cold
protons and relativistically hot electrons. Finally, in
Sec.~\ref{sec:curr-inst} we discuss the possibility of current
instabilities and their role in shaping the precursor. We summarize
our results and draw conclusions in Sec.~\ref{sec:disc}. In
Appendix~\ref{sec:beam-chi}, we explicit the susceptibility tensor of
the beam of finite angular dispersion, modeled with a waterbag
distribution in the transverse momentum direction (transverse with
respect to the shock normal) and a Dirac distribution in the parallel
momentum direction.

\section{General considerations}\label{sec:prelim}

\subsection{Shock structure and precursor}\label{sec:prelim-struct}
Let us first describe the general structure of a relativistic
collisionless shock and introduce the main quantities of interest for
the present discussion.

The shock is characterized by a small number of parameters: (i) the
composition ahead of the shock front, in the upstream plasma -- in
what follows, we assume except otherwise noted that the plasma is
composed of electrons and protons with densities $n_e = n_p \equiv
n_{\rm u}$; (ii) the equation of state of the upstream plasma, which
will be taken cold or composed of relativistically hot electrons but
cold protons (see also further below for more details on this point);
(iii) the shock velocity, written as $\beta_{\rm sh}c$ as measured in
the upstream frame, with corresponding Lorentz factor $\gamma_{\rm
  sh}$; (iv) the magnetization of the upstream plasma, written
$\sigma_{\rm u}$ and defined as
\begin{equation}
\sigma_{\rm u}\,\equiv\, \frac{B_{\rm u}^2}{4\pi n_{\rm u} m_p c^2} \
.
\end{equation}
The magnetic field strength $B_{\rm u}$ is here measured in the
upstream frame.  The shock structure also depends on the obliquity of
the background magnetic field, but in what follows and except
otherwise noted, we will focus on the generic superluminal case in
which the angle with respect to the shock normal exceeds
$1/\gamma_{\rm sh}$.  Then, in the shock front rest frame, the
magnetic field lies mostly perpendicular to the flow direction, as the
transverse components are amplified by a factor $\gamma_{\rm sh}$
while the parallel component remains unchanged by the Lorentz
transform. For an oblique shock wave, the above magnetization
parameter thus measures (up to a $\sin^2\Theta_B\sim 1$ factor, with
$\Theta_B$ the angle between the shock normal and the magnetic field
in the upstream rest frame) the ratio of the magnetic energy in the
shock front frame relative to the incoming matter flux $\gamma_{\rm
  sh}^2n_{\rm u} m_p c^2$ crossing the shock front.

Collisionless shock waves can be mediated by three generic types of
mechanisms: by an electrostatic potential for incoming protons in an
$e-p$ shock, by a magnetic barrier or, if the magnetic field is
negligible, by a reflection due to the ponderomotive force of growing
waves ahead of the shock transition layer. In the case of the external
shock of a gamma-ray burst outflow propagating in the interstellar
medium, for example, the reflection can take place on a potential
barrier or through a ponderomotive force, as the ambient magnetic
field is then particularly weak, $\sigma_{\rm u}\sim 10^{-9}$ for
$B_{\rm u}\sim 1\,\mu$G and $n_{\rm u}\sim 1\,$cm$^{-3}$. However,
even weak, the ambient magnetic field can play an important role, as
will be seen further on.

The magnetic field at a relativistic shock front is associated with a
motional electric field $E_{\rm u\vert sh} = \beta_{\rm sh} B_{\rm
  u\vert sh}$ when measured in the shock front frame (as indicated by
the $_{\vert\rm sh}$ subscript), with $B_{\rm u\vert sh} \simeq
\gamma_{\rm sh} B_{\rm u}$ (again, up to a factor
$\sin\Theta_B$). Since the magnetic field is frozen in most parts of
the plasma, the transverse component is further increased by the
velocity decrease as the upstream incoming plasma approaches the shock
transition layer, as seen again in the shock front rest frame.  This
magnetic barrier may reflect back a fraction of the incoming
protons. In the strongly magnetized case, $\sigma_{\rm u}\gtrsim 0.1$,
the coherent gyration of the incoming protons and electrons gives rise
to the emission of large amplitude electromagnetic waves through the
synchrotron maser instability (Langdon et al. 1988). As these waves
travel back upstream, they lead to a form of wakefield acceleration of
the electrons, which leads to electron heating at the expense of the
incoming protons (Lyubarsky 2006, Hoshino 2008).

The rise of an electrostatic barrier can be described as follows. To
reach thermalization through the shock transition layer, the electrons
need to absorb part of the wave energy, possibly through a kind of
anomalous Joules heating. Whatever the mechanism, the length required
for the electron heating is several inertial length
$\delta_e\,\equiv\,c/\omega_{\rm pe}$, much smaller than the length
scale of the precursor. As electrons remain approximatively in
Boltzmann equilibrium, their density increase in the shock transition
by a compression factor $r$ is accompanied by a potential variation
$\Delta \Phi$ such that $e\Delta \Phi \sim T_e {\rm ln} \, r$. If the
electrons are heated to equipartition with the protons in the
downstream plasma, their temperature $T_e \sim T_p \sim \gamma_{\rm
  sh} m_pc^2$ in the downstream plasma.  In an electron-proton shock,
it thus seems unavoidable to have a potential barrier that reflects
part of the incoming protons (see also Gedalin et al. 2008); this is
an important difference with an electron positron plasma.

If intense waves are excited in the precursor by the reflected
particles or by emitted electromagnetic waves, their growth may also
reflect part of the particles, notably electrons. In particular, such
a mechanism is warranted to ensure the shock transition in an
unmagnetized pair plasma, as evidenced by various PIC simulations, see
for instance Spitkovsky (2008b).

The composition of the beam of returning particles is an important
issue with respect to the development of instabilities in the upstream
plasma, in particular whether charge neutralization has been achieved
or not. In that respect, it should be noted that, if electrons are
heated to near equipartition with the incoming protons in the shock
precursor, the shock itself must behave as if it were a pair
shock. Then, the electrostatic barrier can be expected to be
negligible and the returning beam is essentially charge neutralized.

Whether and to what amount the electrons are heated in the shock
precursor depends in turn on the instabilities that are generated in
this precursor. Let us denote by $\chi_e\leq1$ the fraction of the
incoming (proton) energy carried by the electrons as they reach the
shock transition layer. Then, $\chi_e = m_e/m_p$ if the electrons have
not been heated in the precursor, and $\chi_e = 1$ if equipartition
with the ions has been reached. This parameter $\chi_e$ is now
measured by massive PIC simulations (Spitkovsky 2008a, Sironi \&
Spitkovsky 2009, 2011): generally speaking, it is found that $\chi_e
\sim 1$ for strongly magnetized plasmas, $\chi_e < 1$ in the
intermediate magnetization regime $\sigma_{\rm u}\sim
10^{-4}-10^{-2}$, and $\chi_e$ rises up to unity again at lower
magnetizations.

As discussed in Lemoine \& Pelletier (2010), the returning protons and
the first generation of accelerated particles form a forward beam of
energy $\gamma_{\rm sh}^2m_p c^2$. In the following, we write
$\gamma_{\rm b}=\gamma_{\rm sh}^2$ the Lorentz factor of the protons
composing the beam. If the electrons composing the beam have been
heated to equipartition with the protons before being reinjected
towards upstream, either through reflection or through shock crossing
from downstream toward upstream, they carry a same energy $\gamma_{\rm
  sh}^2m_p c^2$ hence their beam plasma frequency is also similar.
 
In the precursor, the trajectory of a reflected proton is a cycloid
with an extension upstream that determines the width of the ``foot"
region (i.e. the precursor) $\ell_{\rm f\vert sh} = \gamma_{\rm sh}
m_pc^2/ eB_{\rm u\vert sh}$ (measured in the shock front rest
frame). As measured in the upstream frame, this length scale
$\ell_{\rm f\vert u} = \ell_{\rm f\vert sh}/\gamma_{\rm sh}$; this
corresponds to $r_{\rm L\vert u} / \gamma_{\rm sh}^3$ with a Larmor
radius $r_{\rm L\vert u} = \gamma_{\rm sh}^2 m_p c^2/eB_{\rm u}$; the
factor $\gamma_{\rm sh}^3$ comes from the fact that the reflected
protons are caught up by the shock front after having travelled along
the Larmor circle only $r_{\rm L\vert u}/\gamma_{\rm sh}$ before being
caught back by the shock front (Gallant \& Achterberg 1999, Achterberg
et al. 2001), keeping in mind that the distance between the particle
and the front is reduced by a factor $1-\beta_{\rm sh} \simeq
1/(2\gamma_{\rm sh}^2)$ (Milosavljevi\'c \& Nakar 2006, Pelletier et
al. 2009).

When the Fermi process develops through cycles of particles crossing
the front back and forth, the first cycle always occurs and the
particles participating in this first cycle have an energy comparable
to that of the reflected protons. One can thus assimilate the
reflected protons to those first cycle particles, and the number of
particles involved in the Fermi process will only slightly increase
with further cycles. We will note $\xi_{\rm b}$ the ratio of the
incoming energy density converted into the pressure of these
supra-thermal particles:
\begin{equation}
\label{eq:defxi}
\xi_{\rm b}\,\equiv\, \frac{n_{\rm b}m_p
  c^2}{\gamma_{\rm sh}n_{\rm u}m_pc^2}\ ,
\end{equation}
and $n_{\rm b}$ denotes the proper density of the beam particles
(i.e., measured in the shock front rest frame). This parameter
$\xi_{\rm b}$ approximately corresponds to the ratio of the
supra-thermal particle density over the upstream particle density
measured in this same shock rest frame.

\subsection{Instabilities at a relativistic collisionless shock}\label{sec:intro-inst}

In the case of a current neutralized charged beam, meaning a beam
carrying positive and negative charges but zero net current along the
shock normal, the leading instabilities at small magnetization are the
filamentation mode, the two-stream instabilities, in particular the
\v{C}erenkov resonance mode with electrostatic modes (or with modes of
wavenumber parallel to the magnetic field in the case of a magnetized
plasma) or with Whistler modes (for an electron-proton plasma). The
filamentation instability takes place at small real frequencies ${\cal
  R}\omega\sim0$ and small parallel wavenumber
$k_\parallel\,\ll\,k_\perp$ relatively to the transverse wavenumbers
(parallel and tranverse are defined relatively to the shock normal),
while the two-stream electrostatic instability takes place at
resonance ${\cal R}\omega\simeq\omega_{\rm p}$, $k_\parallel \simeq
\omega_{\rm p}/c$ in the upstream rest frame, with $\omega_{\rm
  p}=(4\pi n_{\rm u}e^2/m_e)^{1/2}$ the background plasma frequency.

Such charge driven current neutralized instabilities are of course
generic in the case of pair shocks. However, even in that case, one
must expect a transverse current to rise at the tip of the precursor,
as a consequence of charge splitting in the external magnetic
field. This current then generates a compensating current in the
background plasma, which induces a Buneman instability, which itself
leads to efficient heating of the background electrons in the shock
precursor. This issue will be addressed in Sec.~\ref{sec:curr-inst}.

In the case of an electron-proton plasma, the beam can be current
neutralized if, as mentioned above, the incoming upstream electrons
are preheated to near equipartition with the protons during the
crossing of the precursor; then, for all practical matters, the shock
transition resembles that of a pair shock and the returning beam
carries a vanishing current. Of course, in that case as well, one
expects the transverse current to emerge in a magnetized upstream
plasma and give rise to a Buneman type instability, see
Sec.~\ref{sec:curr-inst}. As mentioned before, such near equipartition
has been observed in PIC simulations in both the high and low
magnetization limits in oblique collisionless shock waves (Sironi \&
Spitkovsky 2011). At high magnetization, electron heating in the
precursor is due to wakefield acceleration associated to the
ponderomotive force of the large amplitude waves emitted by the
synchrotron maser instability (Lyubarsky 2006, Hoshino 2008). At low
magnetization, electron heating appears to be related to the
development of micro-instabilities in the shock precursor.

Such micro-turbulent heating can be understood as follows. Consider
for simplicity the fully unmagnetized limit and select the upstream
rest frame; assume further that the shock has reached a stationary
state. The micro-turbulence is excited on a length scale $\ell'_{\rm
  f\vert u}$ that differs from the previous value of the foot length
$\ell_{\rm f\vert u}$, as we now neglect any pre-existing magnetic
field and assume that scattering is dominated by the micro-turbulence
(see Milosavlejvi\'c \& Nakar 2006, Pelletier et al. 2009). The
magnitude of this length scale is such that the transverse momenta of
returning particles diffuse by an amount $\langle\Delta p_{\rm
  b,\perp}^2\rangle \sim p_{\rm b}^2/\gamma_{\rm sh}^2$ during their
travel time $2\gamma_{\rm sh}^2 \ell'_{\rm f \vert u}/c$ in this
precursor. The prefactor $2\gamma_{\rm sh}^2$ corresponds, as before,
to the difference between the return timescale of the accelerated
particles and the precursor light crossing time. For
returning/accelerated particles, the momentum $p_{\rm b}\sim
\gamma_{\rm sh}^2m_pc$. Diffusion of momenta takes place through
scatterings on small (plasma) scale electromagnetic fields. Such small
scale fields equally contribute to heating the upstream plasma
electrons, provided that the electric wave energy content is not much
smaller than its magnetic counterpart. However, the upstream electrons
only experience the length scale $\ell'_{\rm f \vert u}$ during a
light crossing time $\ell'_{\rm f \vert u}/c$. Consequently, their
momentum dispersion amounts to $\Delta p_{\rm u}^2 \sim m_p^2 c^2/2$
once the electrons reach the shock front, which corresponds to
equipartition with the incoming ions.

To go one step further, one can easily conceive that the preheating of
the electrons to near equipartition in the precursor provides the only
means to achieve near neutralization of the returning particle
current. If indeed electrons are not preheated in the precursor but
simply overturned (by say, the micro-turbulence), their Lorentz factor
is increased by a factor $\gamma_{\rm sh}^2$ but their energy remains
a factor $m_e/m_p$ below that of the returning proton
beam. Consequently, their penetration length scale is also a factor
$m_e/m_p$ smaller and for all relevant purposes, the beam can be
considered as essentially composed of protons.

In short, a net current along the shock normal in the upstream plasma
may arise if and only if the electrons have not been heated to
equipartition with the incoming ions by the time they reach the shock
front. According to the recent PIC simulations of Sironi \& Spitkovsky
(2011), parallel shock waves offer such an example in which the
electrons reach the shock transition with an energy that is
significantly less than that of the protons. It is then natural to
expect current instabilities, such as the Bell (2004) instability --
more exactly, its relativistic generalization, see Reville et
al. (2006) -- to develop in the upstream plasma (Lemoine \& Pelletier
2010). This will be discussed in Sec.~\ref{sec:curr-inst}.

In the forthcoming Sec.~\ref{sec:coldback} and \ref{sec:hotback}, we
concentrate on the filamentation and two stream instabilites and we
discuss their development once the finite angular dispersion of the
beam of accelerated particles has been taken into account, and
considering the possibility that the electrons of the background
(upstream) plasma have been preheated through the micro-turbulence to
relativistic temperatures. Following the approximation scheme
  discussed in our previous paper (Lemoine \& Pelletier 2010), we
  consider the micro-instabilities triggered in the ambient plasma by
  the returning beam over a length scale that is limited by a low
  background magnetization. In that paper, which discussed the
  instabilities under the assumption of a unidirectional beam and a
  cold ambient plasma, we showed that the Weibel-filamentation
  instability and the oblique two stream instability in particular are
  very weakly modified by the magnetic field during their linear
  growth, notably because the maximum length scale of the precursor is
  always much smaller than the Larmor radius of the beam. Moreover the
  magnetization of the ambient plasma is always assumed much smaller
  than unity.  Thus, in this paper, we neglect the contribution of the
  magnetic field to the dispersion tensor. Of course, such a
  contribution cannot be neglected when discussing the development of
  Whistler waves. However we do not consider such an instability in
  the present paper as one of our conclusions is that upstream
  electrons are heated towards equipartition, which quenches the
  development of Whistler waves.

\section{Instabilities with a cold background
  plasma and finite beam angular dispersion}\label{sec:coldback}

\subsection{Beam geometry}

Henceforth, we consider a current neutralized beam composed of
electrons and protons moving with bulk Lorentz factor $\gamma_{\rm b}$
in the upstream rest frame. In this section, we furthermore assume the
background plasma to be cold.

We model the beam with the following axisymmetric waterbag
distribution function:
\begin{equation}
f_{\rm b}(\mathbf{u})\,=\,\frac{1}{\pi
  u_{\perp}^2}\delta(u_x -
u_{\parallel})\Theta(u_{\perp}^2 - u_y^2-u_z^2)\ ,
\end{equation}
introducing the velocity variables $u_i = p_i/(mc)$, where $p_i$
denotes the $i-$component of the beam momentum.  The beam thus
propagates towards $+x$ (parallel direction) and suffers from angular
dispersion in the perpendicular plane, as characterized by the ratio
$u_\perp/u_\parallel$.

In the case of particles returning from or through a relativistic
shock of Lorentz factor $\gamma_{\rm sh}$ and propagating in the
unshocked plasma, the amount of angular dispersion is known to be
$u_\perp\simeq u_\parallel/\gamma_{\rm sh}$. This result is dictated
by kinematics: given that the shock front is always trailing right
behind the accelerated particle, once the parallel velocity of this
latter drops below $\beta_{\rm sh}$, i.e. once the perpendicular
velocity of this latter exceeds $c/\gamma_{\rm sh}$, the particle is
caught back by the shock wave (Gallant \& Achterberg 1999, Achterberg
et al. 2001).  As the beam (proton) Lorentz factor $\gamma_{\rm
  b}\simeq \gamma_{\rm sh}^2$ and $\gamma_{\rm sh}\gg1$, this implies
$u_{\perp}\,\ll\, u_{\parallel}$, hence
$u_{\parallel}\,\simeq\,\gamma_{\rm b}$ and $u_{\perp}\simeq
\gamma_{\rm sh}\,\simeq\, \sqrt{\gamma_{\rm b}}$. This hierarchy
allows to calculate the susceptibility tensor of the beam by
neglecting in a systematic way $u_y^2$ and $u_z^2$ in front of
$u_x^2$. Of course, one must pay attention not to neglect $u_y,\, u_z$
in the poles of the form $\omega - \mathbf{k}\cdot\bfs{\beta}c\,=\,
\omega - k_i u_i c/\gamma_{\rm b} $. The corresponding beam
susceptibility tensor $\chi^{\rm b}_{ij}$ is detailed in
Appendix~\ref{sec:beam-chi}.

In the final expression for $\chi^{\rm b}_{ij}$, the angular
dispersion enters through the ratio $k_\perp u_\perp c/R_\parallel$,
with $R_\parallel= \gamma_{\rm b}\omega - k_\parallel u_\parallel c$
[see App.~\ref{sec:beam-chi}]. Therefore its impact on the growth of
instabilities is determined by $k_\perp$, $\gamma_{\rm sh}$ and the
nature of the instability which determines $R_\parallel$.  At
\v{C}erenkov resonance, $ R_\parallel \simeq \gamma_{\rm b}\omega_k
\delta$, with $\omega_k$ the eigenmode pulsation, and $\delta$ a
complex number of modulus $\vert\delta\vert\,\ll\,1$; at leading
order, $\delta$ is a cubic root of unity in the limit
$u_\perp\rightarrow 0$ (see Lemoine \& Pelletier 2010), hence the
growth rate ${\cal I}\omega = \omega_k {\cal I}\delta\approx \vert
R_\parallel/\gamma_{\rm b}\vert$ (at resonance). Consequently, the
ratio $\left\vert k_\perp u_\perp c/R_\parallel\right\vert\approx
k_\perp \beta_\perp c/ {\cal I}\omega$. If $k_\perp \beta_\perp c/
{\cal I}\omega>1$, the angular dispersion terms dominate over the
resonance poles, hence one expects growth to be inhibited. Formally,
the beam susceptibility changes structure in this limit, as discussed
in App.~\ref{sec:beam-chi}. Physically, one recovers the argument of
Akhiezer (1975), that the particles travel in the perpendicular
direction more than one wavelength of the unstable mode on an efolding
timescale of the instability, hence coherence is lost and growth
inhibited, as discussed recently by Rabinak et al. (2011).  Another
important instability in the precursor of relativistic shocks is the
filamentation mode, with ${\cal R}\omega \sim 0$, $k_\parallel \,\ll\,
k_\perp$. Then $k_\perp u_\perp c/ R_\parallel \,\sim\,k_\perp
\beta_\perp c/ {\cal I}\omega$ as well.

Assuming, as we do in App.~\ref{sec:beam-chi} that the
transverse component of the wavenumber lies along $\mathbf{y}$, the
dispersion relation to be solved reads
\begin{eqnarray}
\biggl(\omega^2 - \omega_{\rm p}^2 - k_\perp^2 &+& \chi^{\rm
    b}_{xx}\omega^2\biggr)
\left(\omega^2 - \omega_{\rm p}^2 - k_\parallel^2 + \chi^{\rm b}_{yy}\omega^2\right) 
          \nonumber\\
&& -\left(k_\parallel k_\perp + \chi^{\rm
             b}_{xy}\omega^2\right)^2\,=\,0\ .\label{eq:disp-rel}
\end{eqnarray}
The susceptibility tensor $\chi^{\rm b}_{ij}$ scales as the ratio
squared of the beam plasma frequency to the background plasma
frequency, i.e.
\begin{equation}
\left(\frac{\omega_{\rm pb}}{\omega_{\rm p}}\right)^2\,=\, \xi_{\rm
  b}\frac{m_e}{m_p}\, ,\label{eq:omegapb}
\end{equation}
with $\xi_{\rm b}$ defined in Eq.~(\ref{eq:defxi}).  In the following,
we solve numerically the dispersion relation for the growth rates of
these various instability modes and discuss the inhibition of
instabilities due to the angular dispersion of the beam.  In all
numerical calculations below, we fix $\xi_{\rm b}=0.1$.

\subsection{Oblique two stream instability}
The oblique two stream instability corresponds to the \v{C}erenkov
resonance of the relativistic beam with the Langmuir modes of the
background plasma. In the case of a magnetized background plasma, the
electrostatic modes that are excited propagate along the background
magnetic field. Since the treatment is similar, we focus here on the
unmagnetized case.

\begin{figure}
\includegraphics[width=0.49\textwidth]{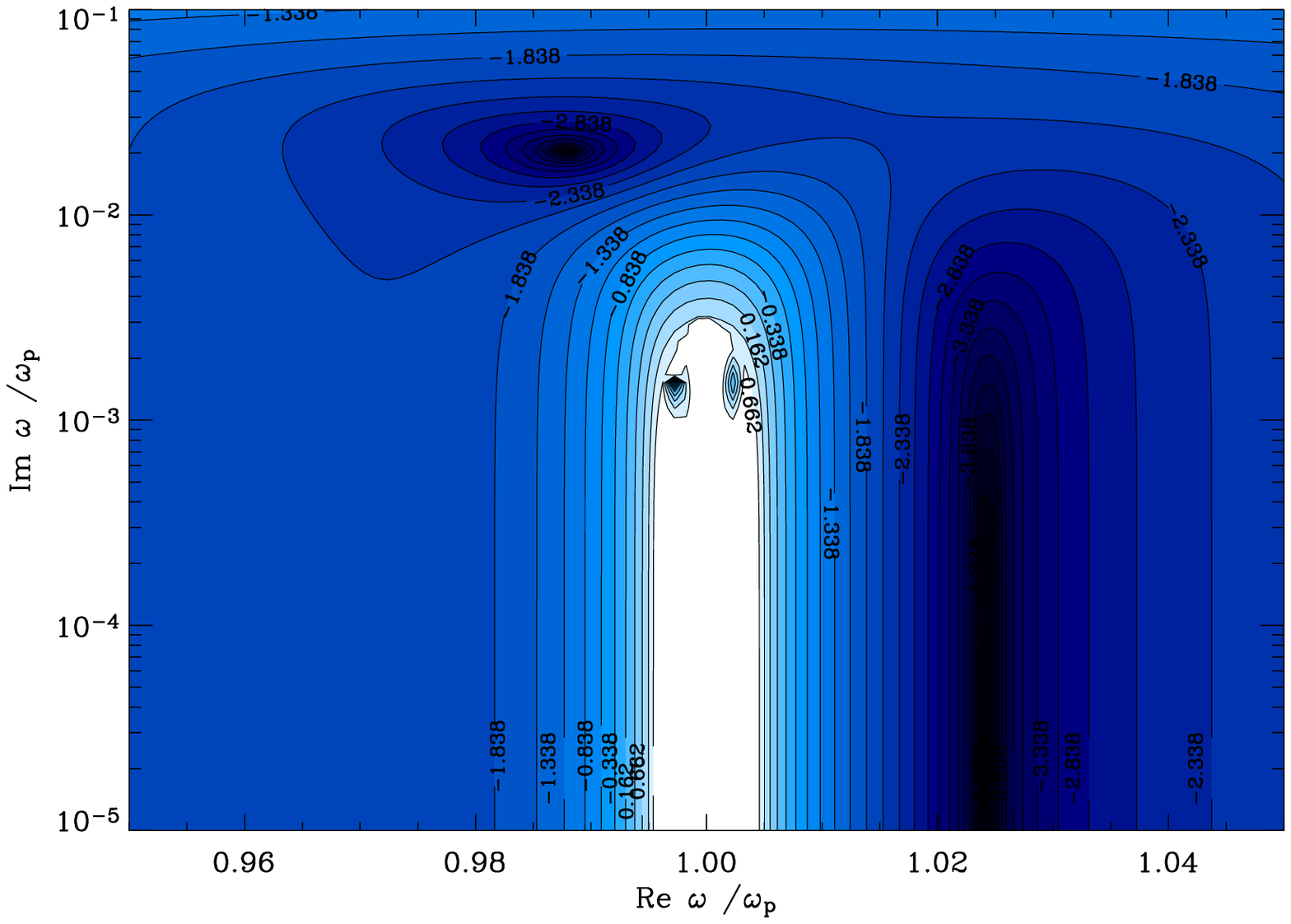}
\includegraphics[width=0.49\textwidth]{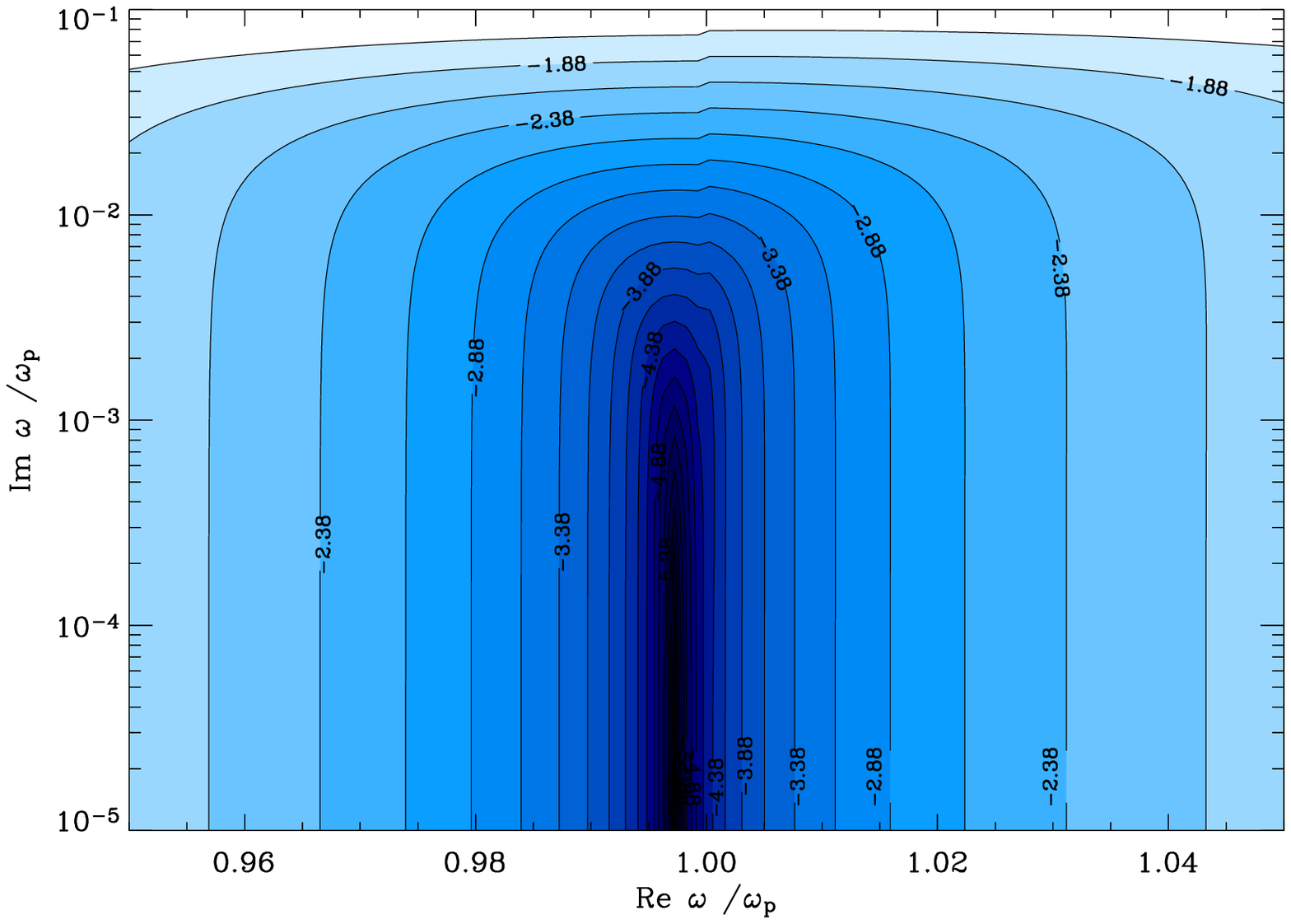}
\caption{Contour map of the log$_{10}$ of the l.h.s. of the dispersion
  relation Eq.~(\ref{eq:disp-rel}) for a cold background plasma,
  including the beam contribution. Top panel: $\gamma_{\rm b}=10^5$,
  the electrostatic eigenmode includes one branch with zero imaginary
  part and one growing mode (as well as one decaying mode, not shown
  here), as indicated by the darker regions. Bottom panel:
  $\gamma_{\rm b}=10^2$, the growing mode has disappeared due to the
  increased angular dispersion of the beam. In both plots, it is
  assumed $\omega_{\rm pb}^2/\omega_{\rm p}^2 = 0.1 m_e/m_p$,
  $k_\parallel = \omega_{\rm p}/(\beta_{\rm b}c)$ (resonance
  condition), $k_\perp=\omega_{\rm p}/c$.\label{fig:disp_OTSI} }
\end{figure}

To evaluate numerically the growth rate, we proceed as follows. We
impose the resonance condition $k_\parallel=\omega_{\rm p}/(\beta_{\rm
  b}c)$ and look for solutions of the full dispersion relation in the
half plane $({\cal R}\omega,\,{\cal I}\omega>0)$, including the beam
contribution, for each given value of $k_\perp$. As the beam slightly
modifies the real part of the root, this latter is slightly displaced
from its unperturbed value $\omega_{\rm
  p}$. Figure~\ref{fig:disp_OTSI} offers an example of this
procedure. It shows the locations of the roots of the dispersion
relation in the $({\cal R}\omega,{\cal I}\omega)$ plane for
$\gamma_{\rm b}=10^5$ (top panel) and $\gamma_{\rm b}=10^2$ (bottom
panel), with $k_\perp =\omega_{\rm p}/c$ in both cases. The growing
mode is clearly seen in the top panel, but absent in the bottom panel
in which the condition $k_\perp \beta_\perp c < {\cal I}\omega$ is
violated due to the smaller value of $\gamma_{\rm b}$.

To leading order in $\chi^{\rm b}$, the growth rate reads (see Lemoine
\& Pelletier 2010)
\begin{equation}
{\cal I}\omega\,\simeq\,\frac{\sqrt{3}}{2^{4/3}}\left(\omega_{\rm
    pb}^2\omega_{\rm p}\right)^{1/3}\left(\frac{k_\perp^2c^2 +
    \omega_{\rm p}^2/\gamma_{\rm b}^2}{k_\perp^2c^2 + \omega_{\rm p}^2}\right)^{1/3}
\ .\label{eq:Iomega_OTSI_cold}
\end{equation}
Therefore, the condition $k_\perp \beta_\perp c \,\ll\, {\cal
  I}\omega$ amounts to 
\begin{equation}
\gamma_{\rm sh}\,\gg\, \xi_{\rm
  b}^{-1/3}\left(\frac{m_e}{m_p}\right)^{-1/3}
\left(\frac{k_\perp c}{\omega_{\rm p}}\right)^{1/3}\quad\quad
\left(\frac{\omega_{\rm p}}{\gamma_{\rm b}}\,\ll\,k_\perp c\,\ll\,
  \omega_{\rm p}\right)\ .\label{eq:otsi-bound}
\end{equation}
One can thus check that, indeed, for $\gamma_{\rm sh}=10$
(corresponding to $\gamma_{\rm b}\simeq 100$), the above condition is
violated at $k_\perp c/\omega_{\rm p}=1$, while it is satisfied for
$\gamma_{\rm sh}=300$ ($\gamma_{\rm b}=10^5$), in nice agreement with
the numerical evaluations.

The effect of the beam temperature on the growth of the two
  stream oblique instability has been discussed previously by Bret et
  al. (2005b) and Bret et al. (2010), although in a slightly different
  context. They show that the two stream instability is rather immune
  to angular dispersion contrary to the filamentation mode, as we find
  here, although the maximum growth rate of the oblique two stream
  mode decreases with increasing dispersion; our findings also match
  these conclusions. In particular, as the angular dispersion
  increases, the maximum growth rate is found at smaller values of
  $k_\perp$, with a reduced growth rate.

\begin{figure}
\centerline{\includegraphics[width=0.49\textwidth]{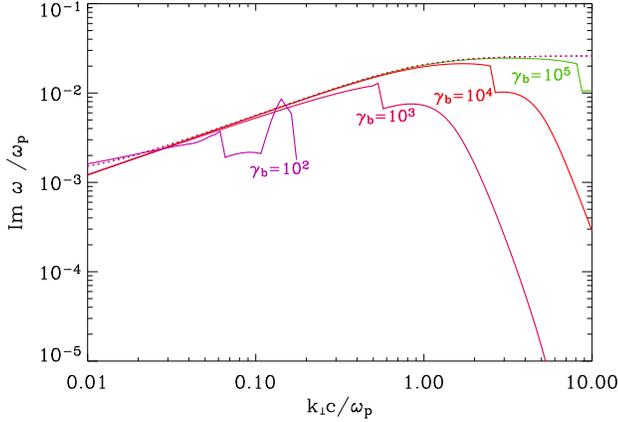}}
\caption{Growth rate of the oblique two stream instability vs $k_\perp
  c/\omega_{\rm p}$ for various values of $\gamma_{\rm b}$: from left
  to right, $\gamma_{\rm b}=10^2,\,10^3,\,10^4,\,10^5$. The cut-offs
  at high frequency are associated to the inhibition of growth due to
  angular dispersion of the beam; $\omega_{\rm pb}^2/\omega_{\rm p}^2
  = 0.1 m_e/m_p$, $k_\parallel = \omega_{\rm p}/(\beta_{\rm b}c)$
  (resonance condition). The dotted line represents the analytical
  solution for the growing mode (see Lemoine \& Pelletier
  2010).\label{fig:cutoff_OTSI} }
\end{figure}

Figure~\ref{fig:cutoff_OTSI} presents the numerical evaluation of the
growth rate as a function of $k_\perp$ (solid line) for various values
of $\gamma_{\rm b}$ (hence, various values of $u_\perp$). This figure
reveals clearly the wavenumber cut-offs that correspond to the finite
beam angular dispersion.

The smallest value of $\gamma_{\rm sh}$ that allows growth of the
oblique two stream instability corresponds to setting $k_\perp
\rightarrow \omega_{\rm p}/\gamma_{\rm b}$ in
Eq.~(\ref{eq:otsi-bound}) above, which leads to
\begin{equation}
\gamma_{\rm sh}\,\gg\, \xi_{\rm b}^{-1/5}\left(\frac{m_e}{m_p}\right)^{-1/5}\ .
\end{equation}
Note that, at values $k_\perp\lesssim \omega_{\rm p}/\gamma_{\rm b}$,
the oblique two stream mode has reduced to the standard parallel
two-stream instability. Furthermore, one must keep in mind that the
above condition dictates whether growth may occur when angular
dispersion is taken into account, yet for growth to occur, other
conditions must be met. Most notably, the growth timescale must be
shorter than the precursor crossing timescale. This latter condition
depends on the degree of magnetization of the ambient medium, as
discussed in detail in Lemoine \& Pelletier (2010).

\subsection{Filamentation instability}
\begin{figure}
\centerline{\includegraphics[width=0.49\textwidth]{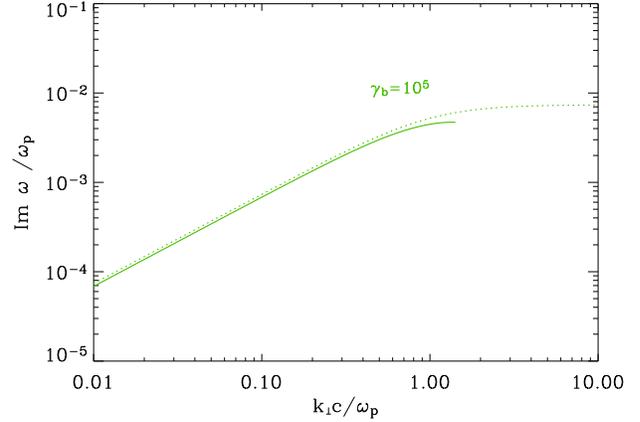}}
\caption{Same as Fig.~\ref{fig:cutoff_OTSI}, but for the filamentation
  instability. The calculation assumes $k_\parallel=0$. Only the
  growth rate for $\gamma_{\rm b}=10^5$ is shown, as smaller values of
  $\gamma_{\rm b}$ did not lead to exact growing
  solutions. \label{fig:cutoff_weibel} }
\end{figure}

The filamentation instability appears at small values of $k_\parallel$
and ${\cal R}\omega$, as the solutions to the dispersion relation
given in Eq.~(\ref{eq:disp-rel}).  The effect of the beam angular
dispersion on the growth rate of the filamentation instability has
been discussed by Bret et al. (2005b), Bret et al. (2010), and in
  the context of relativistic shocks by Lyubarsky \& Eichler (2006)
and recently by Rabinak et al. (2011). This instability turns out to
be extremely sensitive to the beam angular dispersion, and as soon as
$\gamma_{\rm b}\,\lesssim 10^4$, the growth is inhibited for all
values of $k_\perp$. One important difference relative to the oblique
two stream instability is the smaller growth rate of the filamentation
instability, all things being equal, which implies that the condition
$k_\perp \beta_\perp c > {\cal I}\omega$ is more easily satisfied at
given values of $\gamma_{\rm b}$ and $k_\perp$. This is notably
illustrated by Fig.~\ref{fig:cutoff_weibel}, which shows the growth
rate of the filamentation mode for various values of $\gamma_{\rm
  b}$. This numerical calculation considers the limit $k_\parallel
\rightarrow 0$ and finds the roots of the dispersion relation for
various values of $k_\perp$, as before.

In detail, the growth rate of the filamentation instability in a cold
background plasma is given at leading order by
\begin{equation}
{\cal I}\omega\,=\,\omega_{\rm pb}\left(\frac{k_\perp^2c^2}{k_\perp^2c^2 +
    \omega_{\rm p}^2}\right)^{1/2}\ ,
\end{equation}
therefore, ${\cal I}\omega\,\gg\, k_\perp\beta_\perp c$ implies
\begin{equation}
\gamma_{\rm sh}\,\gg\,\xi_{\rm
  b}^{-1/2}\left(\frac{m_e}{m_p}\right)^{-1/2}\quad\quad \left(k_\perp c\,\ll\,
\omega_{\rm p}\right)\ ,\label{eq:cond-fil-cold}
\end{equation}
independently of the value of $k_\perp$ as long as $k_\perp
c\,\ll\,\omega_{\rm p}$. At larger values of $k_\perp$, the right hand
side of of Eq.~(\ref{eq:cond-fil-cold}) must be multiplied by $k_\perp
c/\omega_{\rm p}$ hence the condition is more stringent.

For $\xi_{\rm b}=0.1$, the numerical solution confirms that the
filamentation instability disappears as soon as $\gamma_{\rm
  b}\,\lesssim\,10^4$, while it exists at wavenumbers shorter than
$\omega_{\rm p}/c$ for $\gamma_{\rm b}=10^5$. This agrees well with
the above discussion.

\section{Instabilities in a hot background plasma with a finite beam
  angular dispersion}\label{sec:hotback}
As discussed in some length in Sec.~\ref{sec:prelim}, one expects the
incoming electrons to be heated to relativistic energies as they
approach the shock front. Most notably, this preheating has been
observed in various configurations in the latest PIC simulations of
Sironi \& Spitkovsky (2011). In particular, at low
magnetization, the incoming upstream electrons carry about as much
energy as the ions when they cross the shock transition. If half of
the incoming proton kinetic energy density is transferred to the
electron component and both electrons and protons retain a same bulk
Lorentz factor, then one can show easily that the electrons are heated
to a thermal Lorentz factor $\gamma_e \sim m_p/m_e$ in the electron
fluid rest frame.

We thus consider here a hot electron background plasma, in the
ultra-relativistic limit; we assume that the protons remain cold.  We
also retain the unmagnetized background plasma approximation and
assume a ultra-relativistic Maxwellian distribution,
\begin{equation}
f_e(\mathbf{p}) \,=\, n_e\frac{c^3}{8\pi k_{\rm B}^3T_e^3}e^{-p c/k_{\rm
    B}T_e} \ .
\end{equation}
The longitudinal and transverse (with respect to $\mathbf{k}$)
permittivities then read (e.g. Silin 1960; see also Hakim \& Mangeney
1971, Melrose 1982, Braaten \& Segel 1993, Bergman \& Eliasson 2001)
\begin{eqnarray}
\epsilon^{\rm L}&\,=\,& 1 + \frac{\mu \omega_{\rm p}^2}{k^2c^2}\left[1
+ \frac{\omega}{2kc}\,{\rm ln}\left(\frac{\omega-kc}{\omega+kc}\right)\right]
\label{eq:epsilonL_hot}\ , \\
\epsilon^{\rm T}&\,=\,& 1 - \frac{\mu \omega_{\rm p}^2}{2k^2c^2}\left[1
+ \frac{\omega^2 - k^2 c^2}{2\omega\,kc}\,{\rm ln}
\left(\frac{\omega-kc}{\omega+kc}\right)\right]\label{eq:epsilonT_hot}\ ,
\end{eqnarray}
with $\gamma_e = 3/\mu$, $\mu=m_e c^2/(k_{\rm B}T_e)$.

In what follows, the relativistic background plasma frequency is
written $\Omega_{\rm p}\,\equiv\, \omega_{\rm p}/\sqrt{\gamma_e}$.

\subsection{Two stream instability}
In the fully relativistic regime, the longitudinal mode has a
refractive index smaller than unity for all values of $k$, hence the
\v{C}erenkov resonance with the unperturbed eigenmode can never be
fully satisfied. In all rigor, one should derive the dispersion tensor
allowing for corrections of order in $m_ec^2/k_{\rm B}T$, as one is
seeking a resonance at a refractive index of order $1/\beta_{\rm
  b}\simeq 1 + 1/(2\gamma_{\rm b})^2$ with a background plasma in
which the mean Lorentz factor $\gamma_e$ is much smaller than
$\gamma_{\rm b}$. Such a dispersion tensor has been proposed by
Braaten \& Segel (1993). It takes the same form as that given in
Eqs.~(\ref{eq:epsilonL_hot}), (\ref{eq:epsilonT_hot}) although all $k
c$ must be replaced by $k v_*$, where $v_*$ is an effective thermal
velocity of electrons in the background plasma; hence $v_*\simeq
\beta_e c$. Then, one finds that the refractive index of the
longitudinal mode becomes larger than unity (i.e. the dispersion
relation crosses the light cone) at some value
$k_*\simeq\sqrt{3}\left({\rm ln}(4\gamma_e)-2\right)^{1/2}\omega_{\rm
  p}/c$. For $k\gtrsim k_*$, resonance is then possible. However,
extrapolating the form of the dispersion tensor to the light-like
region, one finds that the growth rate of the \v{C}erenkov resonant
mode with the unperturbed eigenmode is exponentially suppressed. We
thus ignore this branch in the following.

Once the beam contribution to the dispersion relation is taken into
account, one finds that an approximate resonance takes place for
$\omega\,\simeq\,\Omega_{\rm p}$, $k_\parallel = \omega/ \beta_{\rm b}
c$ for small values $k_\perp \lesssim \Omega_{\rm p}/c$. This
resonance solves exactly the dispersion relation and leads to growth
of the modes. Neglecting the effect of angular dispersion, this can be
understood analytically as follows.

The full dispersion relation reads
\begin{equation}
\left(\Lambda_{xx} + \chi_{xx}^{\rm b}\right)\left(\Lambda_{yy} +
  \chi_{yy}^{\rm b}\right) - \left(\Lambda_{xy} + \chi_{yy}^{\rm
    b}\right)^2\,=\,0\ ,\label{eq:disp-hot}
\end{equation}
with
\begin{eqnarray}
\Lambda_{xx}&\,=\,& \frac{k_\parallel^2}{k^2}\epsilon^{\rm L} + 
\frac{k_\perp^2}{k^2}\epsilon^{\rm T} - \frac{k_\perp^2
  c^2}{\omega^2}\ ,\\
\Lambda_{yy} &\,=\,& \frac{k_\perp^2}{k^2}\epsilon^{\rm L} + 
\frac{k_\parallel^2}{k^2}\epsilon^{\rm T} - \frac{k_\parallel^2
  c^2}{\omega^2}\ ,\\
\Lambda_{xy} &\,=\,& \frac{k_\parallel k_\perp}{k^2}\left(\epsilon^{\rm L} - 
\epsilon^{\rm T} + \frac{k^2c^2}{\omega^2}\right)\ .
\end{eqnarray}

Consider now the limit $k_\perp \,\ll \Omega_{\rm p}/c$, keeping in
mind that resonance implies $k_\parallel \simeq \Omega_{\rm p}/c$. The
dispersion relation then boils down to that of the longitudinal mode with
\begin{equation}
\epsilon^{\rm L} + \chi_{xx}^{\rm b}\,\simeq\,0\ .
\end{equation}
We now use the form of the dispersion tensor proposed by Braaten \&
Segel (1993) and set $\omega = \beta_{\rm p} k_\parallel c(1+\delta)$,
with $\vert \delta \vert \,\ll\,1$ as is customary. In this limit, the
longitudinal response reduces to
\begin{equation}
\epsilon^{\rm L} \,\simeq \, 4 + \frac{3}{2}{\rm
  ln}\left(\frac{1}{4\gamma_e^2} +
  \frac{\delta-\epsilon_\perp}{2}\right)\ ,
\end{equation}
where
$\epsilon_\perp\,\equiv\,(k-k_\parallel)/k_\parallel\,=\,k_\perp^2/2k_\parallel^2\,\ll\,1$.
Then the dispersion relation is solved provided
\begin{equation}
\delta^2\,=\,\frac{\omega_{\rm pb}^2}{\Omega_{\rm
    p}^2}\left(\frac{1}{\gamma_e^2} + \frac{k_\perp^2 c^2}{\Omega_{\rm
      p}^2}\right)\frac{1}{\epsilon^{\rm L}}\ .
\end{equation}
Due to the smallness of the argument of the log in $\epsilon^{\rm L}$,
the real part of $\epsilon^{\rm L}$ is negative, hence the dispersion
relation admits growing solutions. The argument of the log depends on
$\delta$, hence the following solution
\begin{equation}
{\cal I}\omega \,\simeq\, 0.3 \frac{\omega_{\rm pb}}{\Omega_{\rm
    p}}\frac{k_\perp c}{\Omega_{\rm p}}\ ,\label{eq:otsi-hot}
\end{equation}
is valid up to a logarithmic correction (also assuming $k_\perp
\,\gg\,\Omega_{\rm p}/\gamma_{\rm b}c$).

In the opposite limit $k_\perp\,\gg\,\Omega_{\rm p}/c$, one finds two
branches: the \v{C}erenkov resonance with $\omega \sim k c$, which is
exponentially suppressed as mentioned above and the continuation of
the above approximate resonance, with $\omega\simeq k_\parallel
c\simeq\Omega_{\rm p}$. This latter, however is nothing but a form of
filamentation instability, since it corresponds to ${\cal
  R}\omega\,\ll\,k_\perp c$ and $k_\parallel \ll k_\perp$. This branch
will therefore be discussed in the section that follows [see in
particular, Eq.~(\ref{eq:fil-hot-sol3})].

Let us discuss now the effect of angular dispersion, focussing on the
\v{C}erenkov resonance at $k_\perp c\,\ll \, \Omega_{\rm p}\simeq{\cal
  R}\omega\simeq k_\parallel c$. Using Eq.~(\ref{eq:otsi-hot}) above,
one finds that angular dispersion can be neglected, i.e.  $k_\perp
\beta_\perp c \,\ll\, {\cal I}\omega$  provided
\begin{equation}
  \gamma_e \gamma_{\rm b}\,\gg\, 100 \xi_{\rm
    b}^{-1}\left(\frac{m_e}{m_p}\right)^{-1}\ ,
  \label{eq:otsi-temp}
\end{equation}
independently of the value of $k_\perp$, as long as $k_\perp
\,\ll\,\Omega_{\rm p}/c$. This inequality involves both the thermal
Lorentz factor $\gamma_e$ of the background plasma and the Lorentz
factor of the beam $\gamma_{\rm b}\simeq \gamma_{\rm sh}^2$. Thus,
growth can occur provided the temperature of the background plasma is
sufficiently high; the corresponding threshold temperature scales as
$\gamma_{\rm sh}^{-2}$.

\begin{figure}
\includegraphics[width=0.49\textwidth]{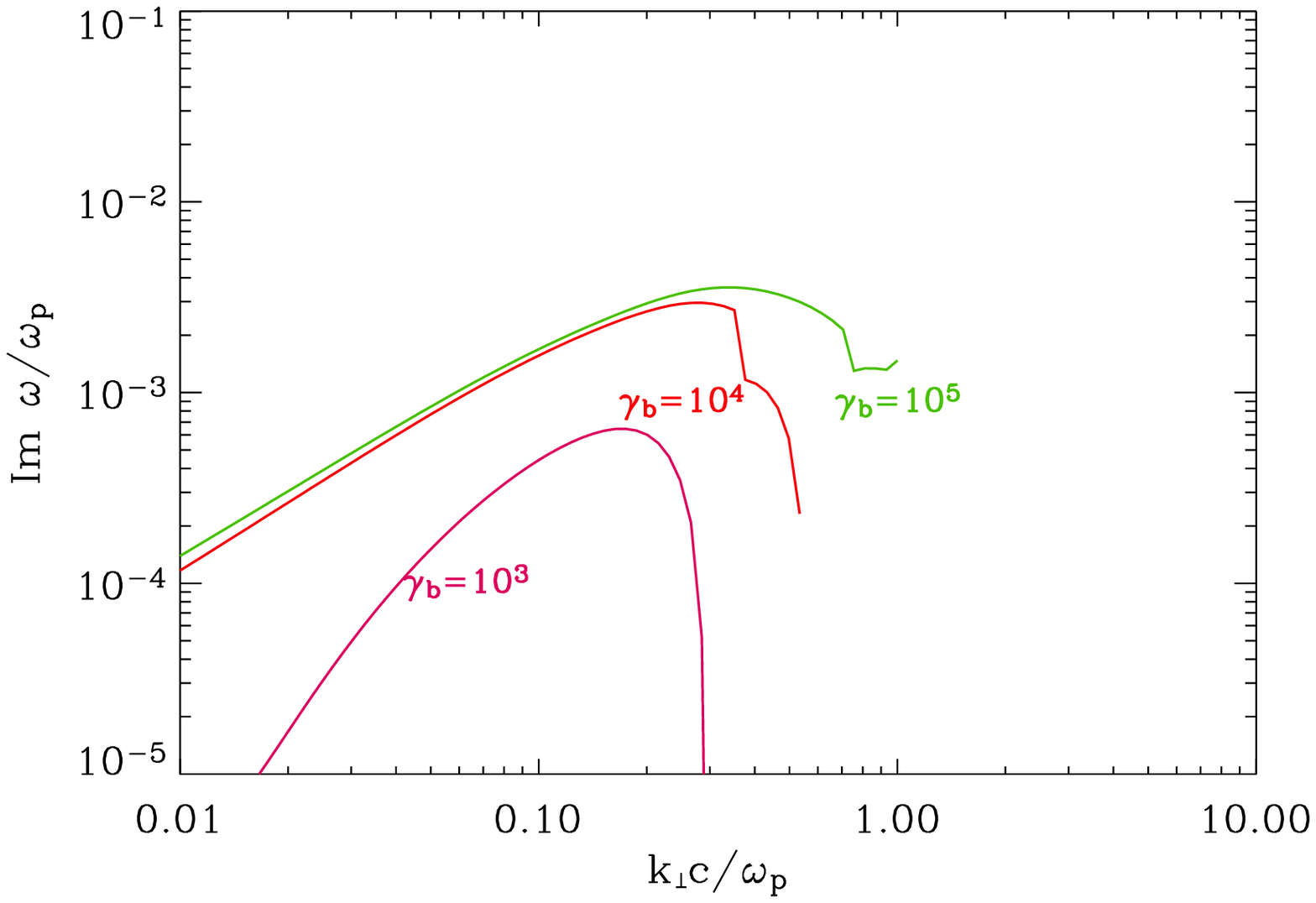}
\includegraphics[width=0.49\textwidth]{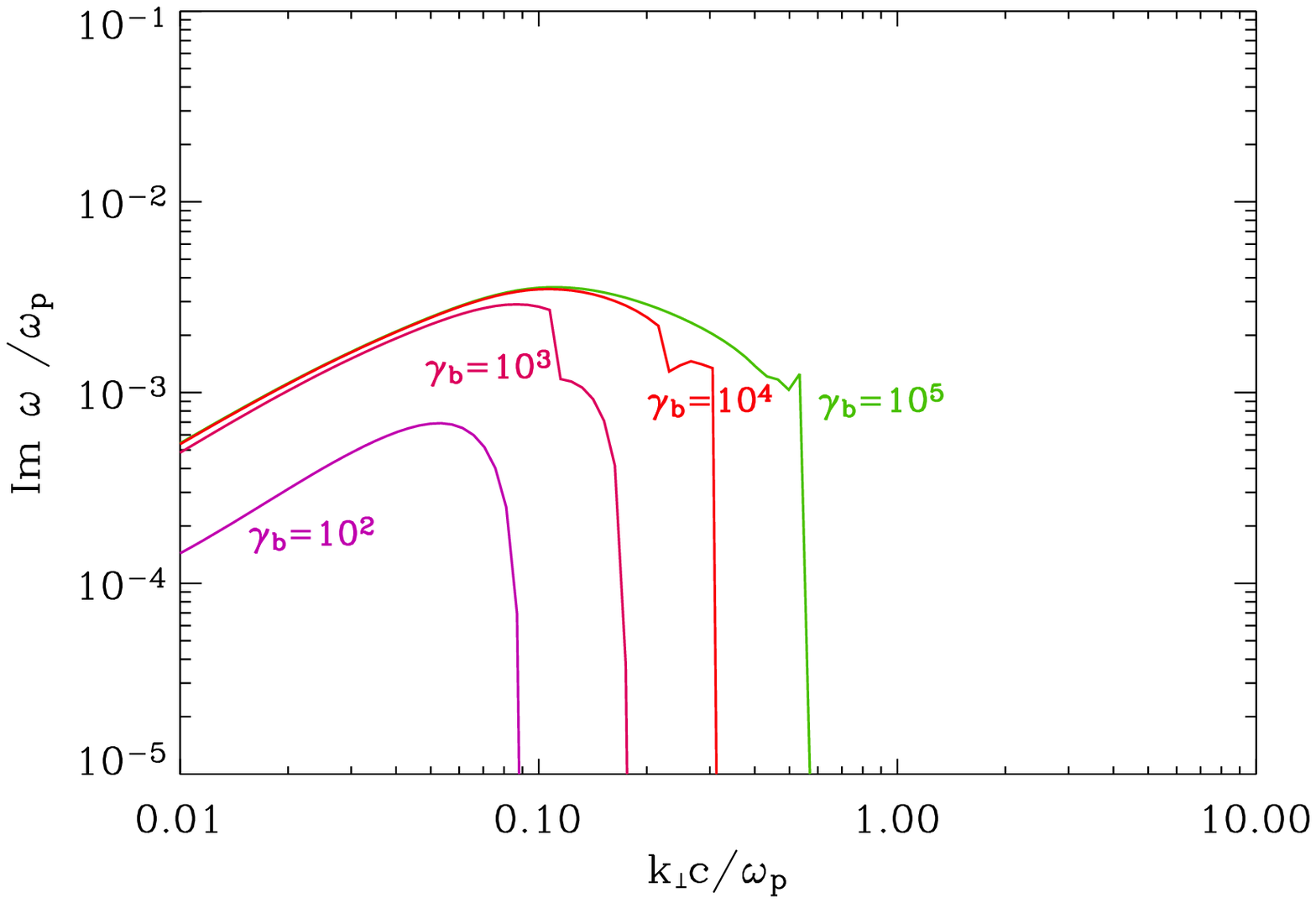}
\caption{Growth rate of the two stream instability in a
  relativistically hot background plasma vs $k_\perp c/\omega_{\rm p}$
  ($\omega_{\rm p}$ non relativistic plasma frequency of the
  background plasma) for various values of $\gamma_{\rm b}$ at mean
  Lorentz factor $\gamma_e=30$ (top panel) and $\gamma_e=300$ (bottom
  panel) of the background plasma electrons. From left to right:
  $\gamma_{\rm b}=10^3$ (bottom panel only), $\gamma_{\rm
    b}=10^4,\,10^5$.  The calculation assumes $k_\parallel c =
  \Omega_{\rm p}/\beta_{\rm b}$. \label{fig:OTSI_T10} }
\end{figure}

Figure~\ref{fig:OTSI_T10} shows the results of a numerical evaluation
of the growth rate as a function of $k_\perp$, for two different
background temperatures, with in each case various values of the beam
Lorentz factor. The agreement with Eq.~(\ref{eq:otsi-temp}) is found
to be quite satisfactory: for $\gamma_e =30$, the instability
disappears at values $\gamma_{\rm b}\,\lesssim\,10^4$, while for
$\gamma_e=300$, it is exists down to values of $\gamma_{\rm b}\sim
10^3$. Note that the growth rate peaks at a value $k_\perp c \sim
\omega_{\rm p}/\sqrt{\gamma_e}=\Omega_{\rm p}$, as one should expect.

\subsection{Filamentation instability}
The filamentation instability in the hot background plasma can be
recovered in the limit ${\cal R}\omega\rightarrow 0$, $k_\parallel
c\rightarrow 0$. We thus write $\omega \equiv iw$, $k_\parallel=0$ and
neglect for the purpose of analytical calculations the angular
dispersion of the beam. This form of instability has been discussed
recently in  Achterberg \& Wiersma (2007). The background plasma
response can be approximated as
\begin{eqnarray}
\epsilon^{\rm L}&\,\simeq\,& 1 + 3\frac{\Omega_{\rm p}^2}{k_\perp^2c^2} +
  3i\frac{w^3\Omega_{\rm p}^2}{k_\perp^5 c^5}\ ,\\
\epsilon^{\rm T}&\,\simeq\,& 1 + \frac{3\pi}{4} \frac{\Omega_{\rm
    p}^2}{k_\perp c w} - \frac{3i}{2}\frac{\Omega_{\rm p}^2
  w}{k_\perp^3 c^3}\ .
\end{eqnarray}
Including the beam contribution, the dispersion relation
Eq.~(\ref{eq:disp-hot}) can be rewritten to leading order as
\begin{equation}
\left(1 + \frac{3\pi}{4} \frac{\Omega_{\rm
    p}^2}{k_\perp c w} + \frac{k_\perp^2c^2}{w^2}\right)
\left(1 + 3\frac{\Omega_{\rm p}^2}{k_\perp^2c^2} \right)
- 3\frac{\omega_{\rm pb}^2}{\Omega_{\rm p}^2}\frac{\Omega_{\rm
    p}^4}{w^4}\,\simeq\,0\ .\label{eq:fil-disp-hot}
\end{equation}
We have implicitly assumed $k_\perp c\,\gg\,w$ in the above
equation. Equation~(\ref{eq:fil-disp-hot}) can be solved in the three
following limits.

If $k_\perp c\,\ll\, w^{1/3}\Omega_{\rm p}^{2/3}$, then $\Omega_{\rm
  p}^2/(k_\perp c w)\,\gg\,k_\perp^2c^2/w^2\,\gg\,1$ and $\Omega_{\rm
  p}^2/k_\perp^2c^2\,\gg\,1$, so that one obtains 
\begin{equation}
{\cal I}\omega\,\simeq\,
\left(\frac{4}{3\pi}\right)^{1/3}\left(\frac{\omega_{\rm
      pb}}{\Omega_{\rm p}}\right)^{2/3} k_\perp c\quad\quad \left(k_\perp c
\,\ll\, \omega_{\rm pb}^{1/3}\Omega_{\rm p}^{2/3}\right)\ .\label{eq:fil-hot-sol1}
\end{equation}
The inequality written in parentheses corresponds to the assumption
$k_\perp c\,\ll\, w^{1/3}\Omega_{\rm p}^{2/3}$. 

If $ \omega_{\rm pb}^{1/3}\Omega_{\rm p}^{2/3}\,\ll\,k_\perp
c\,\ll\,\Omega_{\rm p}$, one rather obtains
\begin{equation}
{\cal I}\omega\,\simeq\, \omega_{\rm pb}\quad \quad
\left(\omega_{\rm pb}^{1/3}\Omega_{\rm p}^{2/3}\,\ll\, k_\perp
  c\,\ll\,\Omega_{\rm p}\right)\ ,\label{eq:fil-hot-sol2}
\end{equation}
which corresponds to the standard filamentation growth rate in a
background plasma at high wavenumbers. The width of the band here is
governed by $(\Omega_{\rm p}/\omega_{\rm pb})^{1/3}\simeq
\gamma_e^{-1/6}\xi_{\rm b}^{-1/6}(m_e/m_p)^{-1/6}$. It should thus not
be much larger than unity.

Finally, if $k_\perp c\,\gg\,\Omega_{\rm p}$, one finds
\begin{equation}
{\cal I}\omega\,\simeq\, \sqrt{3}\frac{\omega_{\rm pb}}{\Omega_{\rm
    p}}\frac{\Omega_{\rm p}}{k_\perp c}\quad\quad
\left(\Omega_{\rm p}\,\ll\, k_\perp c\right)\ .\label{eq:fil-hot-sol3}
\end{equation}

Regarding the effect of angular dispersion, one finds that it can be
safely neglected provided as before, $k_\perp \beta_\perp c \,\ll
{\cal I}\omega$, or
\begin{equation}
  \gamma_e^{1/3}\gamma_{\rm b}^{1/2}\,\gg\,\xi_{\rm b}^{-1/3}
\left(\frac{m_e}{m_p}\right)^{-1/3}\ ,\label{eq:cond-fil-hot-1}
\end{equation}
assuming $k_\perp c\,\ll\,\omega_{\rm pb}^{1/3}\Omega_{\rm
  p}^{2/3}$. 

\begin{figure}
\includegraphics[width=0.49\textwidth]{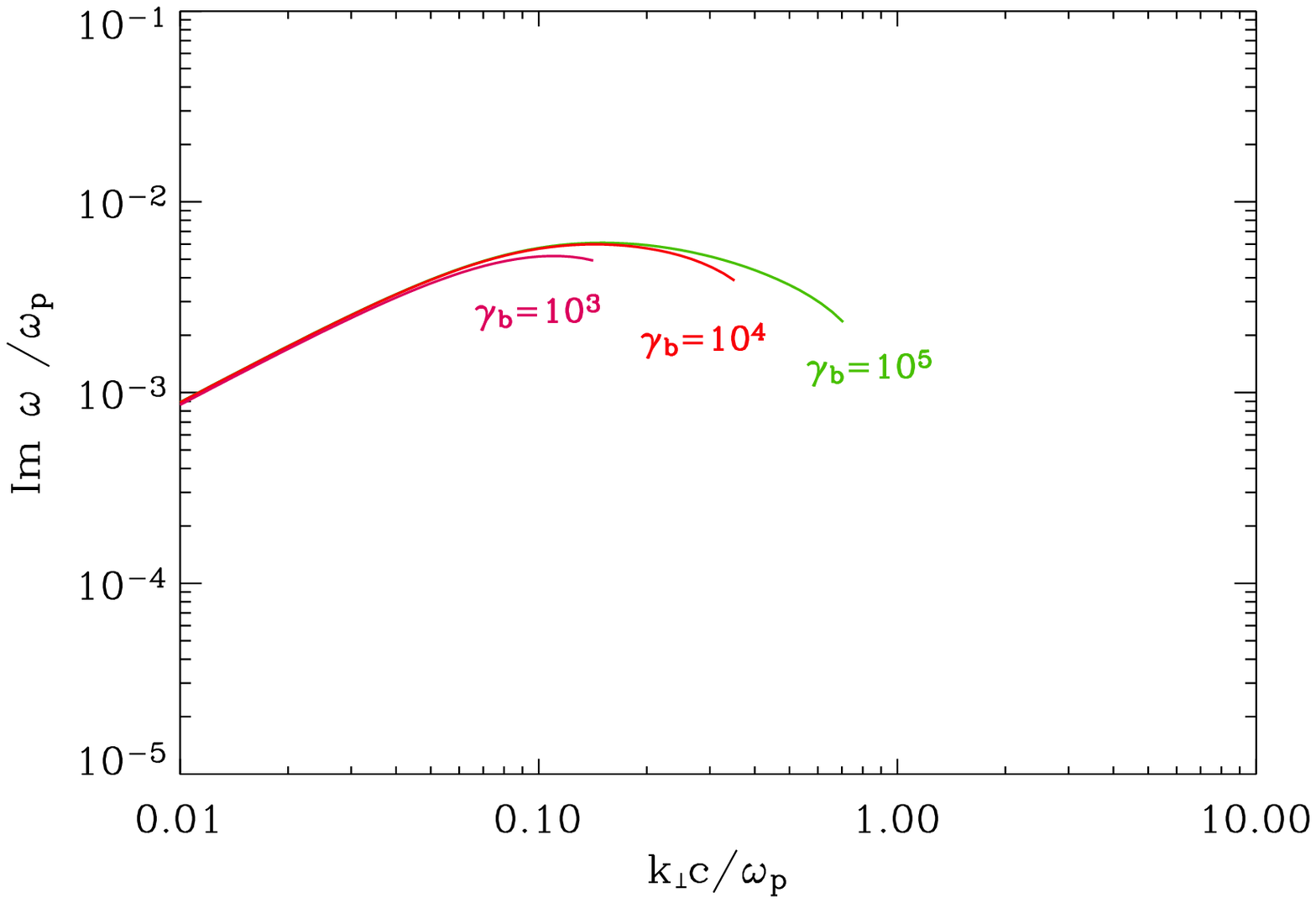}
\includegraphics[width=0.49\textwidth]{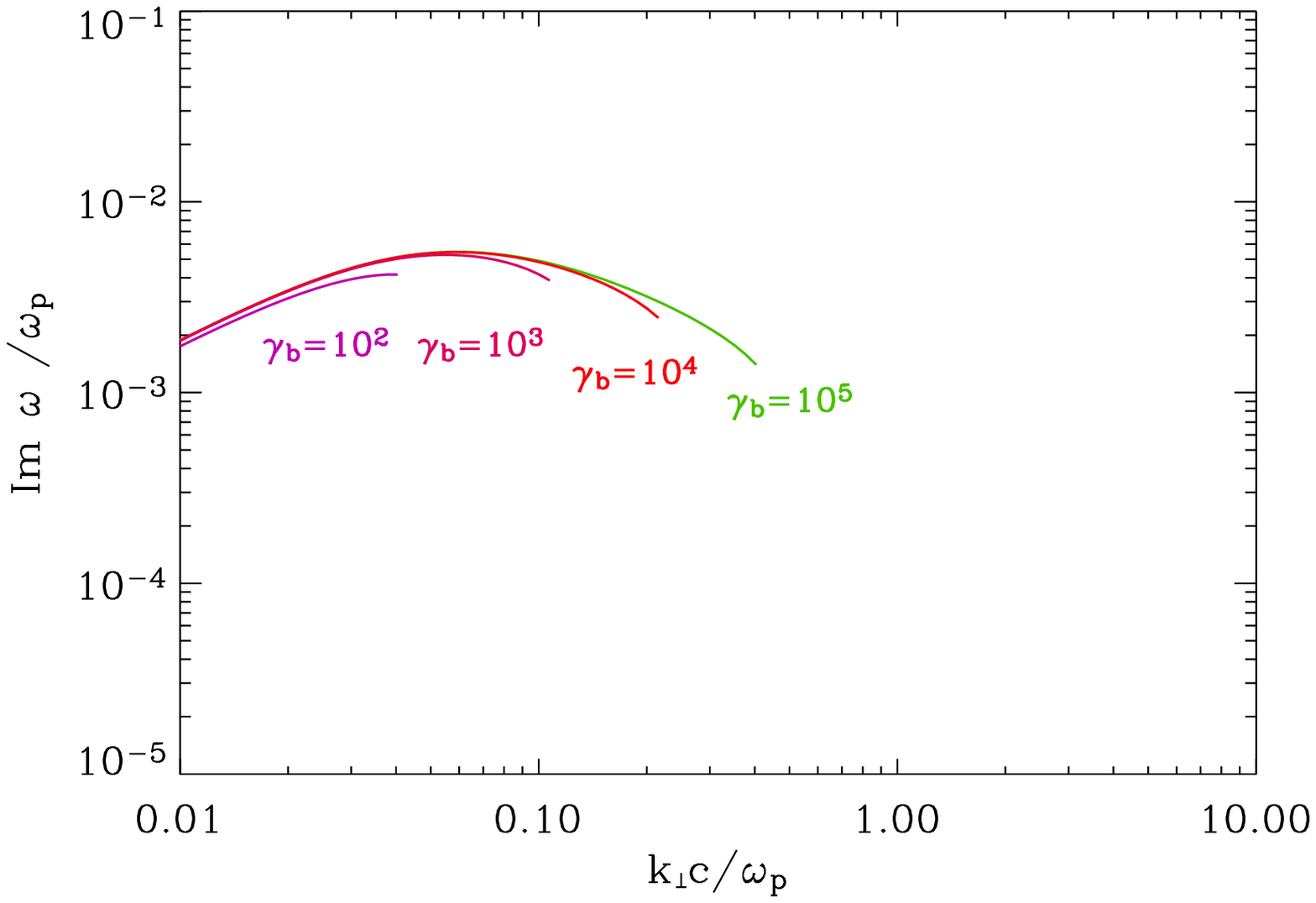}
\caption{Same as Fig.~\ref{fig:OTSI_T10} for the filamentation
  instability. The calculation assumes $k_\parallel=0$. In the upper
  panel, $\gamma_e=30$ while in the lower panel, $\gamma_e=300$. From
  left to right: $\gamma_{\rm b}=10^2$ (lower panel only),
  $\gamma_{\rm b}=10^3,\,10^4,\,10^5$. \label{fig:fil_hot_T10_disp} }
\end{figure}

Figure~\ref{fig:fil_hot_T10_disp} presents the result of a numerical
evaluation of the filamentation growth rate in a relativistically hot
background plasma, for various values of $\gamma_{\rm b}$, taking into
account the angular dispersion. In the left panel, the mean background
Lorentz factor is $\gamma_e=30$ while in the right panel,
$\gamma_e=300$. As expected the growth rate peaks at $k_\perp c \sim
\Omega_{\rm p}$. The angular dispersion shuts off the filamentation
instability at values of $k_\perp$ that scale with $\gamma_{\rm b}$:
indeed, in the range $k_\perp c\,\gg\, \Omega_{\rm p}$, the condition
$k_\perp \beta_\perp c\,\ll\,{\cal I}\omega$ is much more restrictive,
as the right hand side of Eq.~(\ref{eq:cond-fil-hot-1}) is to be
multiplied by $(k_\perp c/\Omega_{\rm p})^2$.

However, contrarily to the results for a cold background plasma shown
in Fig.~\ref{fig:cutoff_weibel}, the growth rate at small wavenumbers
is not strongly affected by the angular dispersion. In this way, the
relativistic temperature of the background sustains the filamentation
instability at small wavenumbers (relatively to $\Omega_{\rm p}/c$)
and moderate values of the beam Lorentz factor.

\subsection{Summary}\label{sec:summ}
Before turning to the current driven instabilities, it may prove
useful to summarize briefly the above results. We have discussed how
the filamentation and the oblique two stream instabilities are
affected by the finite angular dispersion of the beam and the
relativistic temperature of the electrons of the background plasma. In
accordance with previous works (Lyubarsky \& Eichler 2006, Rabinak et
al. 2011), we find that the filamentation instability is strongly
sensitive to the angular dispersion in a cold background plasma. For
an electron-proton plasma, a Lorentz factor $\gamma_{\rm sh}\,\lesssim
100$ shuts off the filamentation instability in a cold background
plasma due to the finite angular dispersion, the extent of which
varies in inverse proportion to $\gamma_{\rm sh}$. Due to its higher
growth rate, the oblique two stream instability is much less sensitive
to the finite angular dispersion and growth may take place at small
wavenumbers for $\gamma_{\rm sh}$ as small as $10-20$. As we have
stressed, this does not take into account the limitation that is
imposed by the size of the precursor: for the mode to actually grow in
a shock precursor, its growth timescale must also be smaller than the
precursor crossing time. This latter condition is discussed in detail
in Lemoine \& Pelletier (2010) and, as it depends directly on the
level of magnetization of the upstream plasma, we do not discuss it
here for simplicity.

Once the electrons are heated to relativistic temperatures, the
picture becomes susbtantially different. Then, the filamentation mode
becomes rather insensitive to the angular dispersion at small
wavenumbers, while the oblique two stream instability is strongly
inhibited. In both cases, the growth rate of the instability is
maximal at a wavenumber $k_\perp \sim \Omega_{\rm p}/c$, which varies
as the inverse square root of the electron temperature. Conversely, a
larger electron temperature implies a larger spatial scale for the
mode of maximum growth rate.

\section{Current instabilities}\label{sec:curr-inst}
Section~\ref{sec:intro-inst} has discussed the possibility of current
drivent instabilities at relativistic shocks. In particular, we have
argued that a net parallel current in the beam of returning particles
may emerge at oblique shock waves (in the upstream frame) if the
electrons of the upstream plasma are not heated to equipartition with
the protons by the time they are overtaken by the shock
  front. We have also indicated that, independently of charge
neutralization of the returning beam, a net perpendicular current may
rise through charge splitting in an external magnetic field. In the
following, we discuss these possibilities in turn.

\subsection{Parallel current}\label{sec:inst-paral-current}
If the shock foot does not preheat the incoming electrons efficiently,
a parallel current with typical (upstream rest frame) intensity
\begin{equation}
j_\parallel \, \sim\,  \xi_{\rm b} \gamma_{\rm sh}^2 n_{\rm u} e c
\end{equation}
rises in the shock precursor. Even if a substantial fraction of the
incoming electrons were reflected at the shock front, the above
current would rise, simply because these electrons would carry an
energy $\sim m_e/m_p$ that of the reflected protons (in the absence of
preheating), hence their penetration length scale in the upstream
would be very small compared to that of the protons.

This current gives rise to a compensating return current in the
upstream plasma, which may destabilize the pre-existing magnetic field
on large spatial scales, see Reville et al. (2007) for parallel
relativistic shock waves.

It is interesting to note that, for realistic values of $\xi_{\rm b}$
and $\gamma_{\rm sh}$, the current exceeds $n_{\rm u}e c$, hence it
cannot be simply compensated by a non-relativistic drift of the
upstream electrons relatively to the ions.  In order to understand how
compensation takes place, it is instructive to go the shock frame, in
which the parallel current $j_{\parallel\vert\rm sh}=\xi_{\rm
  b}\gamma_{\rm sh} n_{\rm u} e c$ is associated to the charge density
$\xi_{\rm b}\gamma_{\rm sh} n_{\rm u} e$. Note that the parallel
  current can be maintained in the shock front frame only if there is
  a net flow of particles escaping toward upstream; this therefore
  implicitly means a subluminal (parallel) shock
  configuration. Although the shock structure is not strictly speaking
  stationary in this case, we assume here that it evolves on long
  timescales.  In this frame, the incoming electrons and protons
enter the precursor with Lorentz factor $\gamma_{\rm sh}$ and density
$\gamma_{\rm sh}n_{\rm u}$. If the electron fluid is suddenly slowed
by an obstacle from velocity $\beta_{\rm sh}$ down to velocity
$\beta_{ex}$, its shock frame density jumps by a factor $\beta_{\rm
  sh}/\beta_{ex}$ but the electron current $j_{ex} = \gamma_{ex}
\beta_{ex} n_{ex} e c$ -- which is positive as the electrons flow in
the negative $x$ direction -- remains conserved. Therefore, the total
incoming $e+p$ current must remain zero. Given that immediately before
entering the precursor, the incoming $e+p$ current vanishes, while
immediately within the precursor, the cosmic ray current does not
vanish, it is easy to see that one cannot balance the current by
simply slowing down the incoming electrons or protons.

In order to achieve charge and current neutralization within the
precursor, it is actually necessary to reflect part of the
electrons. Let us assume that a fraction $\kappa_e$ is reflected back
towards upstream at entrance into the precursor and that electrons and
protons move with respective velocities $\beta_{ex}$ and $\beta_{px}$
within the precursor. The incoming proton current $j_{px} =
-\gamma_{\rm sh}\beta_{\rm sh} n_{\rm u}ec $, the incoming electron
current $j_{ex} = \gamma_{\rm sh}\beta_{\rm sh}(1-\kappa_e)n_{\rm
  u}ec$, the incoming proton charge $\rho_{px} = \gamma_{px} n_{px} e
= \gamma_{\rm sh} n_{\rm u} e \beta_{\rm sh}/\beta_{px}$ and the
incoming electron charge $\rho_{ex} = - \gamma_{ex} n_{ex}(1-\kappa_e)
e = - \gamma_{\rm sh} n_{\rm u} (1-\kappa_e)e \beta_{\rm
  sh}/\beta_{ex}$. Charge and current neutralization within the
precursor thus requires
\begin{equation}
\kappa_e = \xi_{\rm b}, \quad\quad \beta_{ex} = \frac{\beta_{\rm
    sh}}{\xi_{\rm b} + \beta_{\rm sh}/\beta_{px}}\left(1-\kappa_e\right)\ .
\end{equation}
In this case, most of the compensating current in the background
plasma is carried by the incoming protons, while the repelling and the
slowing down of the electrons ensure charge balance. As the incoming
protons are further decelerated by energy exchange with the incoming
electrons through micro-instabilities, the above relation between
$\beta_{ex}$ and $\beta_{px}$ must remain valid to preserve charge
neutralization.  Of course, $\xi_{\rm b}$ itself varies with location,
decreasing roughly exponentially ahead of the shock front. Then, the
above indicate that electrons are slowed down progressively as they
cross the precursor; in the limit $\beta_{px}\rightarrow \beta_{\rm
  sh}$, $\beta_{ex} \rightarrow (1-2\xi_{\rm b})$, leading to
efficient slowing-down of the electrons.  Clearly, one should seek
this effect in current particle-in-cell simulations.

The PIC simulations of Sironi \& Spitkovsky (2011) indicate that the
generation of turbulence does not lead to efficient preheating of the
electrons in the usptream. One may likely relate this to the magnetic
nature of the modes excited by the Bell instability, which do not
contain significant electric wave energy in the upstream
frame. Nonetheless, the precursor length scale increases in time in
the case of parallel shock waves, because a fraction of the
accelerated particles can propagate to upstream infinity in such a
configuration (Lemoine \& Pelletier 2010). Hence, the structure of the
precursor should itself evolve in time: as the length scale of the
precursor increases, incoming electrons experience heating for a
longer duration and arrive at the shock front with a larger fraction
of the incoming ion energy; if this fraction eventually becomes
comparable to unity, the upstream parallel current disappears and the
Bell instability shuts off. Unfortunately, present PIC simulations
have not been able to explore the evolution of the precursor on such
long timescales.

In the case of oblique shock waves, the simulations of Sironi \&
Spitkovsky (2011) indicate that the preheating of the electrons does
not take place efficiently at intermediate magnetisations
$10^{-4}\,\lesssim \sigma_{\rm u}\,\lesssim 10^{-2}$. There is no
  parallel current in this case in the shock frame, only a net charge
  as discussed in Pelletier et al. (2009). A priori, one might expect
  preheating to occur through a Buneman instability induced by the
  perpendicular current (see further below). However, as will be
shown in the following Sec.~\ref{sec:inst-Buneman}, the magnetization
is so large that the Buneman mode does not have time to grow before it
is advected through the shock front, unless the shock Lorentz factor
takes quite moderate values. Hence, it should not bear a significant
impact on the structure of the precursor.

\subsection{Current induced Buneman instability}\label{sec:inst-Buneman}

Let us assume that the accelerated particle beam carries a net current
and induces a compensating return current in the background
plasma. This situation applies equally well to the case of a parallel
(Sec.~\ref{sec:inst-paral-current}) or a perpendicular current
(Sec.~\ref{sec:inst-perp-current}). The relativistic motion of the
background electrons relative to the background ions then induces a
relativistic version of the Buneman instability (Buneman 1959).

In the upstream rest frame, the relevant dispersion relation,
including the relativistic motion of the background electrons with
velocity $\mathbf{v_0}$ (Lorentz factor $\gamma_0$) but neglecting the
beam contribution, takes the form:
\begin{equation}
1 -\frac{\omega_{\rm pi}^2}{\omega^2} 
- \frac{\omega_{pe}^2}{\gamma_e\left(\omega- \mathbf{k} \cdot
    \mathbf{v_0}\right)^2}
\left[1-\frac{\left(\mathbf{k} \cdot \mathbf{v_0}\right)^2}{k^2c^2}\right] = 0 \ .
\end{equation}
The instability develops at frequencies $\vert \omega\vert \ll \omega_0 \equiv
\mathbf{k} \cdot \mathbf{v_0}$, as the small denominator lies in the
second term, not the third one.  Thus we may approximate the
dispersion relation as
\begin{equation}
  1-\frac{m_e}{m_p}\frac{\omega_{\rm pe}^2}{\omega^2} - 
  \frac{\omega_{\rm pe}^2}{\omega_0^2}\frac{k_{\rm tr}^2}{k^2}
  \left(1+2\frac{\omega}{\omega_0}\right)  \simeq 0 \ ,
\end{equation}
where $k_{\rm tr} = \left[k^2 - \left(\mathbf{k} \cdot
    \mathbf{v_0}\right)^2\right]^{1/2}$ represents the component of
$\mathbf{k}$ transverse to the current. The above expression then
leads to the most unstable mode:
\begin{eqnarray}
 \omega_k & \,\simeq\, &
  \left(\frac{m_e}{2m_p}\right)^{1/3}\gamma_0^{-1/6}\left(\frac{k_{\rm
        tr}}{k}\right)^{1/3}\left(-\frac{1}{2} + i \frac{\sqrt
      3}{2}\right) \omega_{\rm p}\ .\\
  \omega_0 &\,\equiv\,&  \mathbf{k}\cdot \mathbf{v_0}  =
  \frac{\omega_{\rm pe}}{\sqrt{\gamma_0}}\frac{k_{\rm tr}}{k} \ .
\end{eqnarray}

As measured in the upstream frame, the advection time across the foot
is $(\gamma_{\rm sh} \omega_{\rm ci})^{-1}$, with $\omega_{\rm ci}=
eB_{\rm u}/(m_pc)$ the proton cyclotron frequency in the background
magnetic field. Thus the Buneman instability can effectively grow if
${\cal I}(\omega_k) \gg \gamma_{sh} \omega_{\rm ci}$ which leads to a
condition on the ambient magnetization, as in {Lemoine \& Pelletier
  (2010)}:
\begin{equation}
\sigma_{\rm u} \ll
\left(\frac{m_p}{m_e}\right)^{1/3}\frac{1}{\gamma_{\rm sh}^2} \ ,
\end{equation}
neglecting the $\gamma_0^{1/6}$ dependence.  This Buneman mode turns
out to be the fastest instability, although the oblique two stream
mode does lie far behind as its growth condition is given by almost
the same inequality, except that the right hand side term is
multiplied by $\xi_b^{1/3}$. Nevertheless, it is well known (at least
in the non-relativistic regime) that this Buneman instability
saturates rapidly by heating the electrons to a temperature such that
the anisotropy due to the electron current is drowned by the broadened
distribution (i.e. $\bar v_e \sim v_e$ in non-relativistic regime,
$\bar v_e$ being the electron thermal velocity, and $\bar \gamma_e
\sim \gamma_e$ in relativistic regime), in agreement with the
discussion of Sec.~\ref{sec:coldback}. In practice, and if conditions
permit it (notably, magnetization), this Buneman mode thus serves as
an efficient source of electron heating. In turn, this helps the
filamentation mode develop at moderate values of $\gamma_{\rm sh}$,
which could not develop in a cold background plasma given the amount
of angular dispersion of the beam, as discussed above.

\subsection{Perpendicular current in a magnetic field}\label{sec:inst-perp-current}

We now consider an oblique shock wave. In the shock front, the
magnetic field lies perpendicular to the shock normal, to a good
approximation; we assume that $\mathbf{B_{\rm u\vert sh}}=B_{\rm
  u\vert sh}\mathbf{y}$. As discussed in Sec.~\ref{sec:prelim-struct},
the returning or acccelerated proton undergoes a cycloidal trajectory
in the Lorentz transformed background field and accompanying motional
electric field, as measured in the shock front frame. Even if the beam
does not carry a net parallel current at the beginning of the foot,
the charge splitting in the background field leads to a cosmic ray
current oriented along $\mathbf{z}$. Its magnitude in the upstream
rest frame can be straightforwardly estimated as $j_{z,\rm b} \sim
\xi_{\rm b}\gamma_{\rm sh} n_{\rm u} e c$, given that the apparent
density of cosmic rays in this rest frame reads $\xi_{\rm
  b}\gamma_{\rm sh}^2 n_{\rm u}$ but that their effective
perpendicular velocity $v_z \sim c/\gamma_{\rm sh}$.

\subsubsection{Magnetic field structure in the precursor}

The vertical current tends to strongly modify the initial magnetic
field in a diamagnetic way. Indeed, the choice of orientation of the
field, directed toward $+y$, implies that the vertical current is
directed toward $+z$, so that the magnetic field increases toward
$+x$, from the shock ramp to the external edge of the foot where it
reaches its external value. A consistent solution thus requires a
reduced mean field in the foot. This remains fully compatible with the
shock crossing conditions, which implies an enhancement of the
magnetic field strength downstream, when one realizes that a similar
current develops behind the shock over a distance measured by a
typical Larmor radius. This current develops in an opposite direction
downstream to that upstream and thus leads to a reduction of the
magnetic field from its far downstream (asymptotic) value to the low
value close to the ramp. Note that the cycloidal trajectories upstream
and downstream have different characteristic gyroradii, due to the
compression of the magnetic field.

This diamagnetic effect could be very prohibitive if the perpendicular
current were not compensated by the background electrons. Indeed, in
the absence of compensation, one would find an induced field $B_{\rm
  ind.}$ such that
\begin{equation}
  \frac{B_{\rm ind.}\left(B_{\rm u\vert sh}-B_{\rm ind.}\right)}{4\pi}
  \sim \xi_{\rm b} \gamma_{\rm sh}^2n_{\rm u} m_pc^2 \ .
\end{equation}

The modification of the magnetic profile can be calculated as follows.
First we remark that the motional electric field that compensates
$\beta_x B_y$ is uniform throughout the shock transition because ${\rm
  rot}\, \vec E = 0$, hence $E_z= \beta_{\rm sh} B_{\rm u\vert
  sh}$. Then, in order to characterize the compensation of the
current, we introduce an effective resistivity $\eta$ of the
background plasma electrons. At the tip of the precursor, where the
beam current is the strongest, one cannot exclude that a ``double
layer'' type of structure forms, as in the case of a parallel shock
wave. For the discussion that follows, we ignore this and describe the
current compensation on phenomenological grounds by introducing this
effective resistivity. In the precursor and in the shock ramp, this
resistivity may result from turbulent scattering at frequency $\nu_e$,
with $\eta = 4\pi \nu_e/\omega_{\rm pe}^2$. Ohm's law for the
background plasma then leads to
\begin{equation}
\beta_{\rm sh} B_{\rm
  u\vert sh} + \beta_x B_y = \frac{\eta}{\gamma_{\rm sh}} j_{z,\rm pl} \ ,
\end{equation}
with $j_{z,\rm pl}$ the background plasma compensating current along
$z$.  Now the field variation is produced by both the plasma current
$j_{z,\rm pl}$ and the diamagnetic cosmic ray current $j_{z,\rm b}$:
\begin{equation}
\frac{\partial B_y}{\partial x} = \frac{4\pi}{c}\left(j_{z,\rm pl}+j_{z,\rm
  b}\right) \ .
\end{equation}
Using Ohm's law to relate $j_{z,\rm pl}$ with $B_y$, one thus obtains
the following differential equation that governs the spatial profile
of the magnetic field:
\begin{equation}
  \frac{\ell_{\rm r}}{\gamma_{\rm sh}}\frac{\partial B_y}{\partial x} - 
  \beta_x B_y = \beta_{\rm sh} B_{\rm u\vert sh} + \frac{4\pi
    \ell_{\rm r}}{\gamma_{\rm sh} c} j_{z,\rm b} \ ,
\end{equation}
where $\ell_{\rm r}$ denotes the resistive length, $\ell_{\rm r} \equiv
\eta c /4\pi = \delta_e^2 \nu_s/c$. The boundary conditions are as
follows: for $x \rightarrow -\infty$ (far downstream), $\beta_x
\rightarrow -1/3$, the current vanish and $B_y \rightarrow 3B_{\rm
  u\vert sh}$; for $x \rightarrow +\infty$ (far upstream),
$\beta_x \rightarrow -\beta_{\rm sh} \simeq -1$, the currents vanish
again and $B_y \rightarrow B_{\rm u\vert sh}$.  In the downstream
region in which there is a cosmic ray current $j_{z,\rm b} < 0$, $B_y$
decreases below $3B_{\rm u\vert sh}$; correspondingly, in the
upstream region in which $j_{z,\rm b} > 0$, $B_y$ increases towards
its asymptotic value $B_{\rm u\vert sh}$.

In the situation that we consider, the magnetic field energy is weak and
the incoming plasma is able to provide a compensating current. The
typical length scale over which the compensating current is
established, $\ell_{\rm r}$ is much smaller than the length scale over
which the cosmic ray current is induced $\ell_{\rm f\vert sh}$, hence one
derives the modification of the magnetic field at $x> \ell_r$ as $B_y
= B_{\rm u\vert sh} + \Delta B^+$ with
\begin{equation}
\Delta B^+ \simeq -\frac{4\pi \ell_r}{\gamma_{\rm sh} \beta_{\rm sh} c} j_{z,\rm b} \ ,
\end{equation}
and the compensating current 
\begin{equation}
j_{z,\rm pl} = \frac{c}{4\pi}\frac{\partial B_y}{\partial x} -
  j_{z,\rm b} = -(1+\frac{\ell_r}{\gamma_{\rm sh}}\frac{\partial}{\partial
    x})j_{z,\rm b} \ .
\end{equation}
The discrepancy with respect to neutralization of the diamagnetic
current is expressed by the derivative term in the above expression,
and is of order $\ell_r / \gamma_{\rm sh} \ell_f$. Thus the diamagnetic
current upstream produces a modification of the magnetic field of a
maximum amount given by
\begin{equation}
  \frac{\Delta B^+}{B_{\rm u\vert sh}} \sim \frac{4\pi \ell_r n_{\rm b}
    c}{\gamma_{\rm sh}B_{\rm u\vert sh}} \sim 4\pi
  \xi_{\rm b}\gamma_{\rm sh}^{-1}\frac{\ell_{\rm
      r}}{\delta_e}\sigma_{\rm u}^{-1/2}\left(\frac{m_e}{m_p}\right)^{-1/2} \ .
\end{equation}
The modification thus remains small if
\begin{equation}
\sigma_{\rm u} \gtrsim \xi_{\rm
  b}^2\left(\frac{m_e}{m_p}\right)\gamma_{\rm sh}^{-2} \sim 10^{-10} \ .
\end{equation}
The right hand side is of order $10^{-6}\gamma_{\rm sh}^{-2}$, hence
it should be verified in most relevant cases.

The above transverse current leads to a compensating current in the
background plasma which induces a Buneman instability, as discussed in
Sec.~\ref{sec:inst-Buneman}. As the beam transverse current is
generated at the tip of the precursor, where most of the rotation in
the background field takes place, the Buneman instability effectively
takes place in a cold background plasma. It may then lead to efficient
heating of the electrons, as discussed in Sec.~\ref{sec:inst-Buneman}.

\section{Discussion and conclusions}\label{sec:disc}
The present work has discussed the electromagnetic micro-instabilities
triggered by a beam of shock reflected/accelerated particles
propagating in the unshocked upstream plasma. In particular, it has
taken into account the finite angular dispersion of the beam of
returning particles as well as the possible effects of heating of the
background plasma electrons. Regarding the development of the
filamentation and two stream instabilities, the salient results are
summarized in Sec.~\ref{sec:summ}. Let us here discuss how these
results affect our understanding of the development of instabilities
at ultra-relativistic shock waves, as a function of the shock Lorentz
factor $\gamma_{\rm sh}$.

1) At large values of $\gamma_{\rm sh}\gtrsim 300$, both the
filamentation and modified two stream instabilities can develop in the
cold or relativistically hot background plasma limits. As mentioned
earlier, this statement only considers the effect of angular
dispersion and plasma temperature. For the waves to actually grow, one
must satisfy another condition: the growth timescale must be shorter
than the precursor crossing timescale. This latter condition depends
on the magnetization of the upstream plasma and is discussed in detail
in Lemoine \& Pelletier (2010). In the rest of this discussion, we do
not consider this limitation, which amounts to considering a very
weakly magnetized upstream plasma.

2) At lower values of $\gamma_{\rm sh}$, the filamentation instability
is inhibited by the finite angular dispersion of the beam -- the
extent of which is inversely proportional to $\gamma_{\rm sh}$ -- in
the cold background plasma limit, but not in the relativistically hot
background plasma limit. The oblique two stream instability is
inhibited in the hot background plasma limit, at least up to some
threshold temperature which scales as $\gamma_{\rm sh}^{-2}$, see
Eq.~(\ref{eq:otsi-temp}).

3) We have uncovered a situation that leads to the development of a
Buneman instability at the tip of the precursor, which may efficiently
preheat the electrons to relativistic temperatures. The Buneman
instability is usually triggered by a parallel current. But, for the
generic case of an oblique relativistic shock wave, the rotation of
the beam in the background magnetic field leads to a perpendicular
current of large intensity. This leads to a compensating transverse
current in the background plasma, which in turn leads to the Buneman
instability, possibly in the relativistic regime. As is well known,
this instability saturates through the heating of the electrons such
that the thermal energy becomes comparable with the drift energy,
thereby drowning the anisotropy of the electron distribution. The
growth rate of the Buneman instability is larger than that of the
oblique two stream instability or the filamentation mode in the cold
background zero angular dispersion limit.

4) The above then suggests that efficient preheating of the upstream
electrons may take place through the Buneman instability and through
the two stream instability. As soon as the electrons have been heated
to relativistic temperatures, the filamentation instability becomes
the dominant mode. As indicated by Eq.~(\ref{eq:cond-fil-hot-1}), this
instability can indeed operate down to values $\gamma_{\rm sh}\sim 10$
in the relativistically hot background plasma limit. For smaller
values of $\gamma_{\rm sh}$, the ultra-relativistic shock limit that
we have assumed throughout is no longer applicable. One should expect
different physics to come into play in the mildly relativistic limit.
\\
The upstream electrons are heated to larger thermal Lorentz factors
$\gamma_e$ as they come closer to the shock front, hence the above
picture must pertain up to the shock front. If the electron
temperature reaches the threshold discussed previously, the two stream
instability also becomes efficient. For reference, if the electrons
reach equipartition with the protons, $\gamma_e$ becomes as large as
$m_p/m_e$. At equipartition, the filamentation instability thus
becomes similar to that occuring in a pair plasma.  As for the
Whistler wave instability, it develops only if there is a sufficient
contrast between the electron and proton masses. It thus disappears
when the electron relativistic mass reaches the proton mass and the
modes become right Alfv\'en waves, which do not have time to grow
unless the precursor has been substantially extended through diffusion
(see the discussion in Lemoine \& Pelletier 2010).

5) The typical perpendicular spatial scale at which the filamentation
and two stream instability growth rates reach their maximum scales as
$\sqrt{\gamma_e}$, due to the evolution of the background plasma
frequency in the ultra-relativistic limit. This suggests that the
typical spatial scale of the inhomogeneities increases from the
electron inertial scale $c/\omega_{\rm pe}$ to the proton inertial
scale $c/\omega_{\rm pi}$ as the modes come closer to the shock
front. Note that the growth rate of the filamentation instability does
not depend on the temperature of the background plasma and takes the
same value as in the cold plasma limit.

6) The micro-turbulence that is generated by these instabilities lead
to efficient electron heating. This has been discussed in
Sec.~\ref{sec:intro-inst}, where it has been argued, in particular,
that in the unmagnetized limit, the large scale of the precursor must
guarantee heating of the electrons to equipartition with the protons
if the wave electric energy content is comparable to the magnetic
content. Let us now account for a finite upstream magnetic field,
assuming in particular that this magnetic field sets the length scale
of the precursor. The heating process takes place through the
diffusion of the electron energy in the bath of micro-turbulence: in
fluctuating electric fields $\bar E$ on coherence scales $l_{\rm c}$,
electrons reach a thermal energy $\bar \epsilon_e$ such that $\bar
\epsilon_e^2 \simeq 2 e^2 \bar E^2 l_{\rm c} \ell_{\rm f\vert u}$
after crossing the precursor of length $\ell_{\rm f\vert u}$. The
turbulence is supposed to reach a level such that it contributes to
form the shock, i.e.  $\bar E^2/(4 \pi) = \xi_{E} n_{\rm u}m_p c^2$
with $\xi_E$ being a conversion factor not too far below unity. This
implies a thermal energy $\bar \epsilon_e \sim \xi_E^{1/2} \sigma_{\rm
  u}^{-1/4} \gamma_{\rm sh}^{-1/2} m_pc^2$. For fiducial values $\xi_E
\sim 10^{-2}$, $\sigma_{\rm u} \sim 10^{-9}$, $\gamma_{\rm sh} \sim 300$,
the pre-factor is of order unity. If one imposes furthermore that the
magnetization is small enough to guarantee the growth of the
filamentation mode, one finds $\sigma_{\rm u}< \xi_{\rm b}/\gamma_{\rm
  sh}^2$ (Lemoine \& Pelletier 2010), the electrons are heated to
Lorentz factors $\gamma_e \gtrsim \xi_E^{1/2} \xi_{\rm b}^{-1/4}
m_p/m_e$, close again to equipartition.
\\
Thus the electrons can be heated in the precursor and roughly
thermalized with the protons.

7) The above considerations agree well with recent PIC simulations, at
least where comparison can be made. In particular, electron heating to
near equipartition has been observed at small magnetization by Sironi
\& Spitkovsky (2011), although the detailed physical process that is
responsible for this heating has not been identified in these
simulations. The micro-instabilities have been observed to take place
at moderate values of the shock Lorentz factor $\gamma_{\rm sh}\sim
20$ and small magnetization $\sigma_{\rm u} \lesssim 10^{-4}$
(assuming in these simulations an electron to proton mass ratio of
1/16). Furthermore, the typical scale of the fluctuations apparently
grows from the tip of the precursor to the shock front.

8) Finally, we question the possible nonlinear saturation of the
instabilities in the precursor. First of all, one can show that these
instabilities cannot be saturated by beam particle trapping because
the timescale to cross a coherence scale $l_{\rm c}$ is much less than
the growth timescale of the waves ${\cal I}\omega^{-1}$; said
otherwise, trapping of the beam particles would require a prohibitive
level of turbulence. This can be seen best by going to the rest frame
of the waves in which the electromagnetic fields are static.  In this
rest frame, the particle cross the transverse coherence length $l_{\rm
  c}$ in a timescale $\tau_{\rm nl\vert w} = (p_{\rm b\vert w}l_{\rm
  c}/e\bar E_{\vert\rm w})^{1/2}$, which takes into account the static
force exerted by the fields on the particle in this rest frame. As we
are interested in the transverse dynamics, one can transform the
relevant quantities to the upstream frame as $\tau_{\rm nl\vert
  u}=\gamma_{\rm w\vert u}\tau_{\rm nl\vert w}$, $p_{\rm b\vert
  w}\simeq p_{\rm b}\simeq \gamma_{\rm sh}^2 m_pc $ and $E_{\vert\rm
  w}\simeq \gamma_{\rm w\vert u} \bar E$. Consider now the example of
the filamentation instability, the growth rate of which is $\xi_{\rm
  b}^{1/2} (m_e/m_p)^{1/2} \omega_{\rm p}$. Saturation by trapping
would require $\xi_E \gtrsim \gamma_{\rm w\vert u}^2\xi_{\rm b}^2
\gamma_{\rm sh}^4 (l_{\rm c}/ \delta_p)^2$, which is obviously
prohibitive.  This negative conclusion obviously holds equally well
for the other instabilities that grow faster, in particular the two
stream instability, the Whistler mode and the Buneman instability.
\\
Nonlinear effects related to mode coupling thus appear more
likely. Even when the electrons have turned relativistically hot, the
rate of nonlinear evolution of the unstable modes -- as for instance
coupling of oblique two stream modes with transverse waves or acoustic
waves -- is expected of the order of $\omega_{pe} \bar E^2/(4\pi
\gamma_e m_e c^2)$ according to Zakharov (1972), which leads to
saturation once the growth rate is balanced by the rate of energy
conversion into the other stable modes. For the particular example of
the two stream instability, with growth rate $\xi_{\rm
  b}^{1/3}(m_e/m_p)^{1/3}\omega_{\rm p}$, this leads to $\xi_E \sim
\xi_{\rm b}^{1/3} (m_e/m_p)^{4/3} \gamma_e$, which is therefore not
far below $\xi_{\rm b}$ if the electrons have reached equipartition
with the ions. As a note of caution, one should point out that the
above estimates of saturation ignore a possible bulk Lorentz factor of
the upstream electrons once they have been heated in the
precursor. Such estimates should nevertheless remain correct in the
limit of moderate $\gamma_{\rm sh}$.

The main conclusion of this study of the dispersion and of the thermal
effects on the growth of micro-instabilities in the foot of a
relativistic shock, is that electron pre-heating can occur, and that
this leads to the attenuation of micro-instabilities except the
filamentation instability, which keeps the same growth rate, the
wavelength of the instability peak migrating from the electron to the
proton inertial scale as the electrons are heated to
equipartition. The plasma that reaches the shock ramp is roughly
thermalized and behaves like an electron-positron plasma. Hence the
following question arises: what makes the reflection of a part of the
incoming particles? Under the assumed conditions of a weak
magnetization, such that it allows the growth of micro-instabilities,
especially the filamentation instability, the role of the
ponderomotive force exerted by the growing waves is probably more
important than the electrostatic barrier. This has to be further
investigated with dedicated PIC simulations.

\section*{Acknowledgments}
We acknowledge fruitful discussions with A. Spitkovsky and L. Sironi
during the preparation of this work, as well as with I. Rabinak and
E. Waxman.  One of us, G.P., is very grateful for the hospitality of
the Kavli Institute for Theoretical Physics in Santa Barbara during
the Summer Program of 2009 devoted to these topics; this was indeed a
great opportunity for developing a better understanding that has
influenced this paper and others to come, thanks to intense exchanges
with many colleagues. We acknowledge support from the CNRS PEPS/PTI
Program of the Institute of Physics (INP) and from the GDR PCHE.

\appendix

\section{Beam susceptibility tensor}\label{sec:beam-chi}

The susceptibility tensor is written as (Melrose 1986):
\begin{eqnarray}
  \chi^{\rm b}_{ij}&\,=\,& \frac{4\pi n_{\rm b}e^2}{m \omega^2}\int {\rm
    d}^3u\, \Biggl[ \frac{u_i}{\gamma} 
  \frac{\partial}{\partial u_j}f_{\rm b} \,+\,\nonumber\\
  && \quad\quad\quad\quad\frac{u_i u_j}{\gamma}
  \frac{1}{\gamma\omega - k_m u_m c + i\epsilon} k_l c 
  \frac{\partial}{\partial u_l}f_{\rm b}\Biggr]\ ,\label{eq:beam}
\end{eqnarray}
with $u_i\,\equiv\,p_i/(mc)$, $\gamma \simeq (1+u_x^2)^{1/2}$ in the
above equation and $n_{\rm b}$ the beam density (in the upstream
plasma rest frame).  We recall the axisymmetric waterbag distribution
function adopted here:
\begin{equation}
f_{\rm b}(\mathbf{u})\,=\,\frac{1}{\pi
  u_{\perp}^2}\delta(u_x -
u_{\parallel})\Theta(u_{\perp}^2 - u_y^2-u_z^2)\ .
\end{equation}
The integral in Eq.~(\ref{eq:beam}) can be carried out analytically
under the approximation discussed in Sec.~\ref{sec:coldback}: we
neglect in a systematic way $u_y^2$ and $u_z^2$ in front of $u_x^2$
but we do not to neglect $u_y,\, u_z$ in the poles of the form $\omega
- \mathbf{k}\cdot\bfs{\beta}c\,=\, \omega - k_i u_i c/\gamma_{\rm b}
$. For simplicity, we rotate the perpendicular axes in such a way as
to align the perpendicular component of the wave vector along $y$,
i.e. $\mathbf{k_\perp} = k_\perp\,\mathbf{y}$. One then obtains
\begin{eqnarray}
  \chi^{\rm b}_{xx}& =& \frac{\omega_{\rm pb}^2}{\omega^2}\,\Biggl\{
    -\frac{1}{\gamma_{\rm b}^2} \nonumber\\
&&\quad\quad\quad- 2k_\parallel u_\parallel
    \Biggl[\,\biggl[\biggl(2 - \beta_{\rm b}^2
          \biggr)\frac{1}{R_\parallel}\,+\,
        \frac{u_\parallel}{\gamma_{\rm b}}\frac{S_\parallel}{R_\parallel^2}\biggr]
      \,
{\cal P}_{xx1}(z)\nonumber\\
&&\quad\quad\quad\quad\quad\quad\quad  - \frac{u_\parallel}{\gamma_{\rm b}}\frac{S_\parallel}{R_\parallel^2}
{\cal P}_{xx2}(z)\Biggr]\nonumber\\
&&\quad\quad\quad+2\frac{u_\parallel^2 k_\perp^2c^2}{R_\parallel^2}
{\cal P}_{xx3}(z)\Biggr\}\
, \label{eq:chixx}\\
\chi^{\rm b}_{xy}& =& 
\frac{\omega_{\rm pb}^2}{\omega^2}\,\Biggl\{
-2\frac{k_\parallel}{k_\perp}\,\Biggl[\,\biggl(\frac{1}{\gamma_{\rm b}^2} +
      2\frac{u_\parallel}{\gamma_{\rm b}}\frac{S_\parallel}{R_\parallel}\biggr)
{\cal P}_{xy1}(z) \nonumber\\
&&\quad\quad\quad\quad\quad\quad +\frac{u_\parallel}{\gamma_{\rm b}}\frac{S_\parallel}{R_\parallel}
{\cal P}_{xy2}(z)\,\Biggr]\nonumber\\
&&\quad\quad\quad+2\frac{u_\parallel k_\perp c}{R_\parallel}
{\cal P}_{xy3}(z)\Biggr\}\ ,\label{eq:chixy}\\
\chi^{\rm b}_{yx}&=&\chi^{\rm b}_{xy}\\
\chi^{\rm b}_{xz}&=&\chi^{\rm b}_{xz}=0 \ ,\label{eq:chixz}\\
\chi^{\rm b}_{yy}&=& \frac{\omega_{\rm pb}^2}{\omega^2}\,\Biggl\{
-1 - 2\frac{k_\parallel}{k_\perp} 
\Biggl[\,\biggl(-\frac{u_\parallel}{\gamma_{\rm
          b}^2}\frac{R_\parallel}{k_\perp c}  + 
  3\frac{S_\parallel}{k_\perp c \gamma_{\rm b}}\biggr)
{\cal P}_{yy1}(z) \nonumber\\
&&\quad\quad\quad\quad\quad\quad\quad\quad+\frac{S_\parallel}{k_\perp c \gamma_{\rm b}}
{\cal P}_{yy2}(z)\,\Biggr] \nonumber\\
&&\quad\quad\quad+2{\cal P}_{yy3}(z)\,\Biggr\}\ ,\label{eq:chiyy}\\
\chi^{\rm b}_{zz}&=&\frac{\omega_{\rm pb}^2}{\omega^2}\,\Biggl\{
-1 - 2\frac{k_\parallel}{k_\perp}  
\Biggl[\,\biggl(-\frac{u_\parallel}{\gamma_{\rm
          b}^2}\frac{R_\parallel}{k_\perp c}  + 
  3\frac{S_\parallel}{k_\perp c \gamma_{\rm b}}\biggr)
{\cal P}_{zz1}(z) \nonumber\\
&&\quad\quad\quad\quad\quad\quad\quad\quad-\frac{S_\parallel}{k_\perp c \gamma_{\rm b}}
{\cal P}_{zz2}(z)\,\Biggr]\nonumber\\
 &&\quad\quad\quad+2{\cal P}_{zz3}(z)\,\Biggr\}\
        ,\label{eq:chizz}\\
\chi^{\rm b}_{yz}&=&\chi^{\rm b}_{yz}=0 \ ,\label{eq:chiyz}
\end{eqnarray}
with 
\begin{eqnarray}
z\,&\equiv&\,\frac{k_\perp u_\perp c}{R_\parallel}\ ,\nonumber\\
R_\parallel\,&\equiv&\, \gamma_{\rm b}\omega - k_\parallel
u_\parallel\,c\ ,\nonumber\\
S_\parallel\,&\equiv&\, u_\parallel \omega -
k_\parallel \gamma_{\rm b}\, c\ .
\end{eqnarray}
The beam plasma frequency is defined as usual: $\omega_{\rm
  pb}\,\equiv\,\left[4\pi n_{\rm b}e^2/(\gamma_{\rm
      b}m)\right]^{1/2}$. The beam velocity $\beta_{\rm b}=
u_\parallel/\gamma_{\rm b}$.

\begin{eqnarray}
{\cal P}_{xx1}(z)&=&
z^{-2}\left[1-\left(1-z^2\right)^{1/2}\right]\ ,\nonumber\\
{\cal P}_{xx2}(z)&=& \left(1-z^2\right)^{-1/2}\ ,\nonumber\\
{\cal P}_{xx3}(z)&=&
z^{-2}\left[1-\left(1-z^2\right)^{-1/2}\right]\ ,\nonumber\\
{\cal P}_{xy1}(z)&=&z^{-2}\left[1 -\frac{z^2}{2} -
  \left(1-z^2\right)^{1/2}\right]\ ,\nonumber\\
{\cal P}_{xy2}(z)&=&1-\left(1-z^2\right)^{-1/2}\ ,\nonumber\\
{\cal P}_{xy3}(z)&=&{\cal P}_{xx3}(z)\ ,\nonumber\\
{\cal P}_{yy1}(z)&=&{\cal P}_{xy1}(z)\ ,\nonumber\\
{\cal P}_{yy2}(z)&=&{\cal P}_{xy2}(z)\ ,\nonumber\\
{\cal P}_{yy3}(z)&=&z^{-2}\left[1 + \frac{z^2}{2} -
  \left(1-z^2\right)^{-1/2}\right]\ ,\nonumber\\
{\cal P}_{zz1}(z)&=&\frac{z^{-2}}{3}\left[ -1 + \frac{3z^2}{2} +
  \left(1-z^2\right)^{3/2}\right]\ ,\nonumber\\
{\cal P}_{zz2}(z)&=&-{\cal P}_{xx1}(z)\ ,\nonumber\\
{\cal P}_{zz3}(z)&=&\left[1-\left(1-z^2\right)^{1/2}\right]\ .
\end{eqnarray}

The above are defined for $k_\perp u_\perp c\,<\,\left\vert
  R_\parallel\right\vert$. In the opposite limit, i.e. when the
effects of angular dispersion become substantial, the integral in
Eq.~\ref{eq:beam} contain poles. Using the Plemelj-Sohotsky formula to
evaluate the integrals over these resonance poles, one finds that the
above
Eqs.~(\ref{eq:chixx},\ref{eq:chixy},\ref{eq:chixz},\ref{eq:chiyy},
\ref{eq:chizz},\ref{eq:chiyz}) for the beam susceptibility tensor are
continued to the region $k_\perp u_\perp c\,>\,\left\vert
  R_\parallel\right\vert$ by the substitution
$\left(1-x^2\right)^{1/2}\rightarrow i\,\left(x^2-1\right)^{1/2}$,
with $x=k_\perp u_\perp c/R_\parallel$ in the expressions for the
$\chi^{\rm b}_{ij}$ components.

\end{document}